\documentclass[acmsmall, screen]{acmart}

\AtBeginDocument{%
  \providecommand\BibTeX{{%
    \normalfont B\kern-0.5em{\scshape i\kern-0.25em b}\kern-0.8em\TeX}}}

\setcopyright{acmlicensed} 
\acmJournal{PACMHCI} 
\acmYear{2025} \acmVolume{9} \acmNumber{2} \acmArticle{CSCW178} \acmMonth{4}\acmDOI{10.1145/3711076}

\usepackage{scalerel}
\usepackage{multirow}
\usepackage{enumitem}
\usepackage{listings}
\usepackage{xparse}
\usepackage[utf8]{inputenc}
\usepackage{xcolor}
\usepackage{colortbl}
\usepackage{makecell}
\usepackage[nameinlink]{cleveref}
\usepackage{array}
\usepackage{wrapfig}
\usepackage{lipsum}
\usepackage{comment}
\usepackage[normalem]{ulem}
\usepackage{hyperref}
\usepackage{hyperxmp}
        
\begin{document}

\title[Societal Impact Assessment for Industry Computing Researchers]{Supporting Industry Computing Researchers in Assessing, Articulating, and Addressing the Potential Negative Societal Impact of Their Work}
%

\author{Wesley Hanwen Deng}
\orcid{0000-0003-3375-5285}
\authornote{This work was conducted while the author was an intern at Microsoft Research.}
\email{hanwend@cs.cmu.edu}
\affiliation{%
  \institution{Carnegie Mellon University}
  \streetaddress{5000 Forbes Ave}
  \city{Pittsburgh}
  \state{PA}
  \postcode{15213}
  \country{USA}
}

\author{Solon Barocas}
\orcid{0000-0003-4577-466X}
\authornote{Both authors contributed equally to this research.}
\email{solon@microsoft.com}
\affiliation{
\institution{Microsoft Research}
  \streetaddress{300 Lafayette St.}
  \city{New York}
  \state{NY}
  \postcode{10012}
  \country{USA}
}

\author{Jennifer Wortman Vaughan}
\orcid{0000-0002-7807-2018}
\authornotemark[2]
\email{jenn@microsoft.com}
\affiliation{
\institution{Microsoft Research}
  \streetaddress{300 Lafayette St.}
  \city{New York}
  \state{NY}
  \postcode{10012}
  \country{USA}
}

\renewcommand{\shortauthors}{Deng, Barocas, and Vaughan}

\begin{abstract}
  Recent years have witnessed increasing calls for computing researchers to grapple with the societal impacts of their work. Tools such as impact assessments have gained prominence as a method to uncover potential impacts, and a number of publication venues now encourage authors to include an impact statement in their submissions. Despite this recent push, little is known about the way researchers go about assessing, articulating, and addressing the potential negative societal impact of their work --- especially in industry settings, where research outcomes are often quickly integrated into products and services. In addition, while there are nascent efforts to support researchers in this task, there remains a dearth of empirically-informed tools and processes. Through interviews with 25 industry computing researchers across different companies and research areas, we identify \textbf{four key factors} that influence how they grapple with (or choose not to grapple with) the societal impact of their research: the relationship between industry researchers and product teams; organizational dynamics and cultures that prioritize innovation and speed; misconceptions about societal impact; and a lack of sufficient infrastructure to support researchers. To develop an effective impact assessment template tailored to industry computing researchers' needs, we conduct an iterative co-design process with these 25 industry researchers, along with an additional 16 researchers and practitioners with prior experience and expertise in reviewing and developing impact assessments or responsible computing practices more broadly. Through the co-design process, we develop \textbf{10 design considerations} to facilitate the effective design, implementation, and adaptation of an impact assessment template for use in industry research settings and beyond, as well as \textbf{our own \href{https://perma.cc/B396-DMTV}{\uline{``Societal Impact Assessment'' template}}} with concrete scaffolds. We explore the effectiveness of this template through a user study with 15 industry research interns, revealing both its strengths and limitations. Finally, we discuss the implications for future researchers, organizations, and policymakers seeking to foster more responsible research practices.
\end{abstract}



\maketitle

\section{Introduction}

In recent years, many organizations across the academy \cite{kieslich2023anticipating, hecht2021s, bernstein2021ethics, Neurips2020workshop, Neurips2020blog, CVPR2023EthicsGuidelines, ACL2023ethicspolicy, ICML2023EthicsGuidelines, olteanu2023responsible, ESR_Stanford}, government \cite{AIA_Adalove, national2022fostering, NAIRRTF2023FinalReport, NAIRRTF2023Strengthening}, civil society \cite{PAI2021managing, ada2022looking, AIA_Adalove, AIML_Data_Society, metcalf2021algorithmic, reisman2018algorithmic}, and industry \cite{RAIIAguide_MSFT, RAIIAtemplate_MSFT, googleRAI, openAI_research, hecht2021s} have published reports and explored tools and processes to better support computing researchers in grappling with the potential negative societal impact of their research. Among other tools, impact assessments have become popular as a way to support both researchers and practitioners
in assessing, articulating, and addressing the impact of their work
\cite{AIA_Adalove, AIA_Canada, AIA_CIO, metcalf2021algorithmic, moss2021assembling}. At a high level, an impact assessment is a systematic process used to identify, assess, and address the potential impact of a proposed project, program, or policy on people, organizations, and society as a whole \cite{PIA_GDPR, RAIIAtemplate_MSFT, metcalf2021algorithmic}.\looseness=-1

Despite these efforts, the path forward remains unclear \cite{boyarskaya2020overcoming,liu2022examining,nanayakkara2021unpacking, ashurst2022ai, ada2022looking, do2023s, benotti2022ethics, selbst2021institutional, johnson2023assessing}. A small but growing line of work has begun to empirically investigate computing researchers' practices around attempting to confront the negative societal impact of their work. This research has revealed that current efforts are not fully effective~\cite{boyarskaya2020overcoming, nanayakkara2021unpacking,liu2022examining, ashurst2022ai}, due in part to a general lack of incentives and support for self-reflection, including from academic institutions \cite{ada2022looking, do2023s, bernstein2021ethics}.

To date, very little of this work has focused on the way \textit{\textbf{industry computing researchers}} in particular go about assessing and addressing the potential negative societal impacts of their work.  Despite a growing line of research on building tools to support practitioners (e.g., product teams) in reflecting and anticipating the societal impact of their work \cite{ballard2019judgment, elsayed2023responsible, wang2024farsight, wong2021timelines, nathan2008envisioning}, prior work from CSCW and the broader field of HCI suggests that resources for responsible computing in industry are not as readily available as one might hope \cite{holstein2019co, rakova2021responsible, deng2022exploring, wong2021tactics, madaio2022assessing, heger2022understanding, wang2023designing}, raising the prospect that industry researchers are similarly lacking in support. Given the proximity that industry research has to practice --- and thus its increased likelihood of being integrated into products and services --- there is an acute need to better understand how industry researchers are currently approaching this challenge. This need is especially urgent in light of the recent rise of generative AI and the risks it poses~\cite{bender2021stochastic,Weidinger2022TaxonomyOR,kumar23language}, since much of the research on generative AI takes place in industry contexts.
\looseness=-1

In addition, while several tools and processes have been proposed to help computing researchers better grapple with the potential negative societal impacts of their work \cite{bernstein2021ethics, AIA_Adalove, AIA_Canada, AIA_CIO, PAI2021managing, NAIRRTF2023FinalReport, reisman2018algorithmic}, few of these are informed by \textit{\textbf{empirical}} research. As HCI research has repeatedly demonstrated, such support is unlikely to be effective unless it is informed by an understanding of the on-the-ground challenges faced by its intended users~\cite{madaio2020co, deng2022exploring, Wong2022SeeingLA, balayn2023fairness, lee2021landscape, richardson2021towards, kaur2020interpreting, yildirim2023investigating}. 

Prior work has pointed out that fostering more responsible research practices in computing  will require significant changes across a range of institutions and sectors \cite{bernstein2021ethics, AIA_Adalove, Neurips2020blog, ICML2023EthicsGuidelines, do2023s, metcalf2021algorithmic, reisman2018algorithmic, rubambiza2024seam}. While we endorse many of these more ambitious proposals, change can be costly and slow, progressing in fits and starts.  Our goal is thus to develop an effective impact assessment tool that researchers can adopt immediately and that can be adapted to different organizational settings and arrangements as these broader changes in practices, policies, and norms ideally unfold. \looseness=-1

To this end, our paper explores the following research questions: \looseness=-1

\begin{itemize} 

\item \textbf{RQ1}: What are industry computing researchers' \textbf{current perceptions, practices, and challenges} around assessing, articulating, and addressing the potential negative societal impact of their research? 

\item \textbf{RQ2}: What are some\textbf{ design considerations} for designing and implementing an impact assessment template to support industry computing researchers in effectively assessing, articulating, and addressing potential negative societal impacts? \looseness=-1

\item \textbf{RQ3}: How might we fulfill these design considerations to develop an \textbf{effective societal impact assessment template} for industry computing researchers?

\end{itemize} 

To investigate our research questions, we first conducted study sessions with 25 industry computing\footnote{We adopt a broad definition of ``computing.'' We cover all areas traditionally viewed as part of computer science, as well as social science and design research that examines the socio-technical aspects of technology along with research at the intersection of computer science and other fields, like biology.  In adopting this broad definition, our aim is to be inclusive of the types of research that typically occur in industry research labs.}
researchers across different organizations and research areas. These researchers are formally employed by private companies, but also publish and present their research findings and innovations at peer-reviewed conferences or journals in the field of computing. Each study session was broken into two components: a semi-structured interview and a co-design activity. The \textbf{semi-structured interview} was aimed at understanding industry researchers' perceptions of the need to grapple with the potential negative societal impact of their research as well as their current practices for doing so, including the challenges they commonly encounter. The \textbf{co-design activity} with researchers was focused on soliciting their needs for effectively conducting impact assessment through the iterative co-design of a research-specific impact assessment, which we call the Societal Impact Assessment (SIA) template. To complement the feedback from industry computing researchers, who represent the intended users of the template, we also involved 16 researchers and practitioners with prior experience in developing and reviewing impact assessments or responsible computing practices more broadly  (whom we refer to as \textit{``impact assessment experts''} throughout this paper) in the co-design process. These participants represent potential administrators or reviewers of the template and are included 
to ensure that the template fulfills its purpose of effective self reflection.
Finally, to further understand the usability and usefulness of the SIA template, we recruited 15 interns from a range of industry research teams at a large U.S.-based technology company.
\looseness=-1

 \begin{figure*}[t]
  \centering
  \includegraphics[width=1\linewidth]{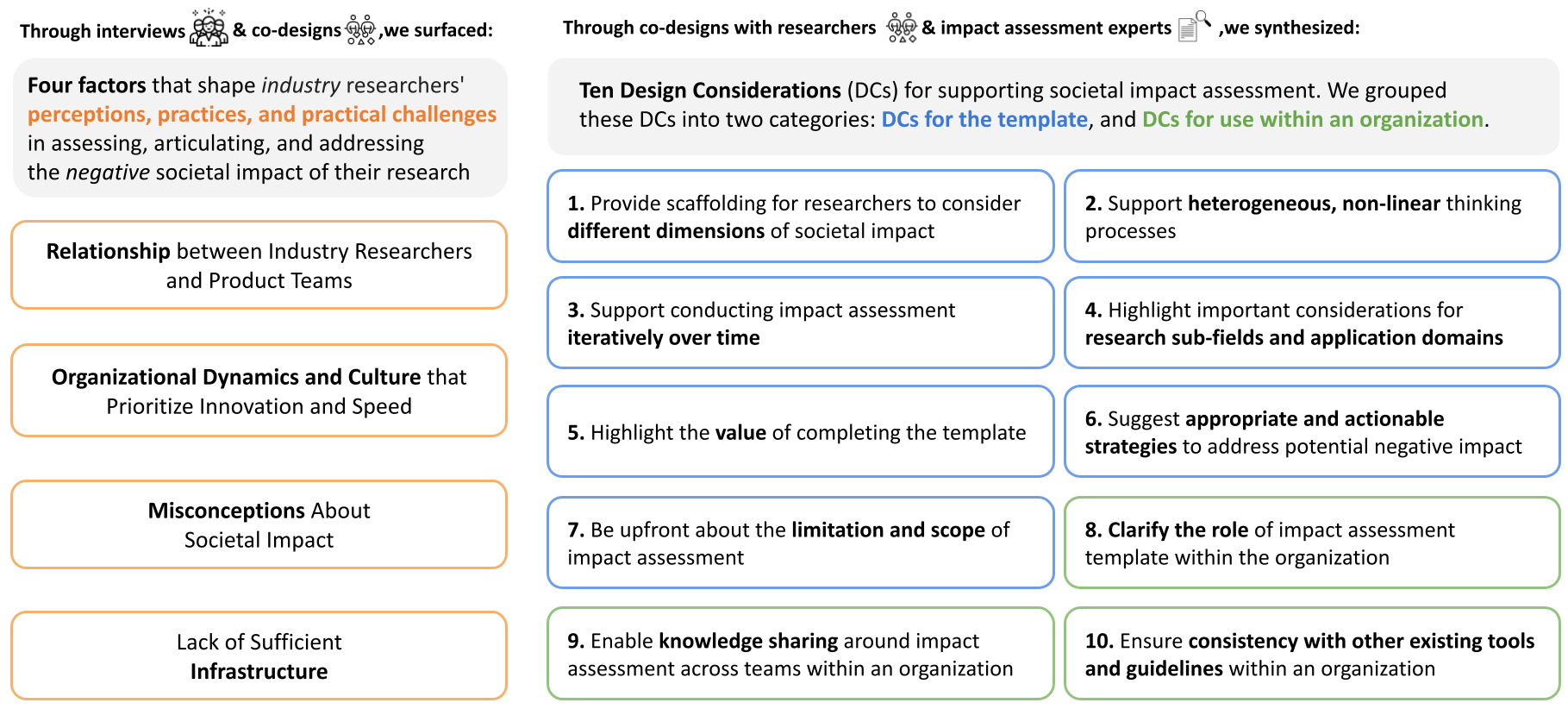}
  \caption{
    An overview of the first two contributions or our work. On the left side, we present the empirical findings on current perceptions, practices, and practical challenges (Section \ref{empirical findings}). On the right side, we present the design considerations for the impact assessment template content and structure (in blue; see Section \ref{DC for template}) and using the template within an organization (in green; see Section \ref{DC for organization}).
}
  \Description{TBA }
  \label{fig:results}
\end{figure*}

Our paper ultimately makes three main contributions:

\begin{itemize}
    
    \item An \textbf{empirical understanding} of industry computing researchers' perceptions, practices, and challenges around assessing, articulating, and addressing the potential negative societal impact of their research. Our findings reveal \textbf{four factors} that influence how industry computing researchers grapple with (or choose not to grapple with) the societal impact of their research: the relationship between industry researchers and product teams (Section \ref{relationship with product}), organizational cultures that prioritize innovation and speed (Section \ref{organizational dynamics}), disagreements and misconceptions about societal impact (Section \ref{misconceptions}), and the lack or inadequacy of infrastructure to support impact assessment in industry settings (Section \ref{lack infrastructure}). \looseness=-1

    \item A set of 10 empirically-informed \textbf{design considerations} to facilitate the proper design, implementation, and adaptation of impact assessment templates in industry research settings and beyond (Section \ref{design considerations}). These design considerations are formulated though the co-design with both industry computing researchers and the impact assessment experts. Figure \ref{fig:results} contains an overview of these design considerations, color-coded to indicate which pertain to template content and structure and which to the use of the template within an organization.

    \item A version of the \textbf{Societal Impact Assessment (SIA) template} iteratively co-designed with industry computing researchers and impact assessment experts across different organizations and research areas (Section \ref{SIA template}), along with results from a \textbf{one-week user study} that reveal the strengths and limitations of the SIA template (Section \ref{user study results}).

\end{itemize}

\section{Background}
Recent calls for computing researchers to grapple with the societal impacts of their work have ushered in a sea-change in thinking about what it means to engage in responsible research in computing. 
Traditionally, the scope of responsible research practices was thought to include threats to research integrity (such as fraud or a failure to disclose limitations) \cite{ashurst2022disentangling, smith2022real} and the dangers posed by the process of conducting research (such as the risks to the human subjects enlisted in a study or the mishandling of their data) \cite{ashurst2022disentangling, bernstein2021ethics}. Recent calls expand the scope to include the risks of downstream harm to society more broadly due to the dissemination of research findings and their associated artifacts. 
Notably, while Institutional Review Boards address ethical issues that arise in the process of conducting research, they purposefully (and are often legally obligated to) avoid considering the dangers that research findings might pose to society overall \cite{united1978belmont, williams2005federal, bernstein2021ethics}. As a result, while there is growing concern around downstream harms, there are no well-established processes for dealing with them. In what follows, we offer an overview of how concerns with societal impact have worked their way through the research community over the past few years and the mechanisms that have been proposed or adopted to address them.
\looseness=-1

In March 2018, Hecht et al. suggested that computer science reviewers should critically examine both the potential positive and negative impacts of research and called for the development of formal guidelines to aid this process \cite{hecht2021s}. In May 2021, the Partnership on AI issued recommendations for the responsible publication of AI research, encouraging researchers to consider the downstream consequences of research early in the research process and disclose relevant information about the risks of downstream harms \cite{PAI2021managing}. The National Academies of Sciences, Engineering, and Medicine followed in May 2022 with a report that likewise called on computer science as a field to integrate concerns with the societal implications of research into the research and publication process \cite{national2022fostering}. In July 2022, the Ada Lovelace Institute, the Canadian Institute For Advanced Research (CIFAR), and the Partnership on AI released a joint report on "A Culture of Ethical AI" offering many similar recommendations targeted at the organizers of AI conferences, with the goal of encouraging greater reflection by researchers on the societal impacts of their work \cite{cifar2022culture}. Later that year, in December 2022, the Ada Lovelace Institute also issued its own, much longer report, offering more detailed recommendations for Ethics Review Committees that could help anticipate and address the societal impacts of AI research \cite{ada2022looking}.

A number of computer science conferences, especially in AI and machine learning, have responded to these calls with changes to their paper submission requirements or instructions and their peer review processes. In 2020, the Neural Information Processing Systems (NeurIPS) conference introduced a requirement that authors ``include a section in their submissions discussing the broader impact of their work, including possible societal consequences --- both positive and negative'' \cite{Neurips2020blog}. A workshop took place at NeurIPS that same year on ``Navigating the Broader Impacts of AI Research'' \cite{Neurips2020workshop}. The subsequent year, in an effort to provide more guidance and flexibility to authors, this requirement was relaxed and a prompt about negative societal impact was incorporated into a broader ``paper checklist'' designed to encourage responsible research practices~\cite{Neurips2021checklist}. Other major AI conferences, such as the International Conference on Machine Learning (ICML), the Conference on Computer Vision and Pattern Recognition (CVPR), and the Annual Meetings of the Association for Computational Linguistics (ACL) have introduced similar requirements \cite{ACL2023ethicspolicy, CVPR2023EthicsGuidelines, ICML2023EthicsGuidelines}, as has the International AAAI Conference on Web and Social Media (ICWSM) \cite{ICWSM}. In 2024, organizers of the ACM Conference on Fairness, Transparency, and Accountability (FAccT) encouraged researchers to include an ``adverse impact statement'' in their submissions, arguing that researchers working on ethical issues in computing need to consider the potential negative societal impact of their own work, too \cite{olteanu2023responsible}. To aid researchers in fulfilling these new requirements or recommendations, conference organizers and researchers in related fields have written informal guides to support computing researchers in writing impact statements \cite{ashurst2020NeurIPS, olteanu2023responsible, hecht2020suggestions, ACL2023ethicspolicy}. \looseness=-1

Finally, funding bodies have also begun to incorporate these considerations into their review processes. In 2020, a cross-disciplinary group of researchers at Stanford developed an Ethics and Society Review Process to ``facilitate ethical and societal reflection as a requirement to access funding,'' focusing on early assessment and mitigation of negative societal impacts in computing research \cite{bernstein2021ethics, ESR_Stanford}. In January 2023, the National AI Research Resource Task Force --- a group convened by the National Science Foundation and White House Office of Science and Technology Policy at the behest of the United States Congress --- issued a report offering recommendations for how the United States government should go about setting up a public resource to support AI research~\cite{NAIRRTF2023FinalReport, NAIRRTF2023Strengthening}. The report suggested that consideration of societal impacts should be a part of the process of reviewing proposals to use public resources. \looseness=-1
\section{Related Work}

\subsection{Challenges in Considering Potential Negative Societal Impact} \label{prior challenges}

Despite the aforementioned growing calls for computing researchers to attend to the potential negative societal impacts of their work, it is still far from clear how this can be done effectively \cite{boyarskaya2020overcoming,liu2022examining,nanayakkara2021unpacking, ashurst2022ai, ada2022looking, do2023s}. For example, examining the broader impact statements written in NeurIPS 2020 papers, Ashurst et al. found that researchers have struggled to fulfill these requirements and to follow these suggestions, and that the quality of the resulting reflections and mitigations is spotty at best \cite{ashurst2022ai}. It appears that the guidelines provided by conference organizers and researchers are not sufficiently effective. Other work also suggested that it is unclear if researchers devoted sufficient effort to reflect on the societal impact of their research beyond simply completing these requirements \cite{nanayakkara2021unpacking, liu2022examining}. 

A small but growing line of empirical work has started to surface practical challenges faced by computing researchers when grappling with the potential negative societal impact of their work \cite{ada2022looking, do2023s, bernstein2021ethics, rubambiza2024seam}. To start with, there are limited tools and systemic guidelines to support assessing, articulating, and addressing potential negative societal impact, leaving researchers to do much of the reflection and possible mitigation on their own \cite{do2023s, bernstein2021ethics}. A report from the Ada Lovelace Institute suggested that traditional research ethics processes, such as IRBs, are often not well-suited for computing research such as AI and ML \cite{ada2022looking}. Moreover, through a series of interviews, Do \& Pang et al. found that academic computing researchers often lack sufficient training, experience, as well as opportunities to collaborate with experts from diverse backgrounds in grappling with unintended consequences of their research \cite{do2023s}. This work suggests that there is a lack of incentives for researchers to consider the negative societal impact of their work, as the process might be at odds with some researchers' goals to publish quickly \cite{do2023s, ada2022looking}. 
In contrast, Rubambiza et al. find that computing researchers may care quite a lot about the societal impact of their work, but are forced to compromise on their vision of societal impact or place less emphasis on societal impact in order to satisfy different audiences (e.g., collaborators, funders, peer reviewers, etc.) whose buy-in is required to execute the research project and publish the findings \cite{rubambiza2024seam}.


Although a small body of prior work has engaged industry computing researchers as part of their study participants in relevant tasks \cite{deng2022exploring, elsayed2023responsible, madaio2024learning}, it does not \textbf{empirically examine the current practices and practical challenges around assessing negative societal impact faced by \textit{industry} researchers}. Understanding these aspects is critical and timely, considering the close relationship between industry research and product and the emerging discussions around the difficulties of assessing impact at later stages of the research-to-practice pipeline \cite{wallach2021navigating}.
Although there is a growing body of CSCW and broader HCI research literature aimed at understanding the current practices, challenges, and needs of industry \textit{developer} teams in building responsible technology \cite{rakova2021responsible,passi2018trust,madaio2020co,madaio2022assessing,heger2022understanding, holstein2019co, wang2023designing, deng2022exploring, widder2023s,  deng2023investigating,  yildirim2023investigating, elsayed2023responsible}, there has been little focus on industry \textit{researchers}. Therefore, our work extends the empirical findings from prior studies to uncover factors influencing how industry computing researchers grapple with (or choose not to grapple with) societal impact. These insights also lay the empirical groundwork for developing tools and processes that support computing researchers in industry and beyond.

\subsection{Tools and Processes to Support Anticipating Potential Harms} \label{existing tools}

CSCW,  design, and HCI research more broadly have a rich history of creating tools and processes aimed at helping designers and developers anticipate potential failures, risks, and harms of the technology they are creating. This work often seeks to foster a greater appreciation of the issues that might arise across various technology deployments and real-world usage scenarios and thus to help developers better anticipate and address them \cite{zhu2018value, steinhardt2015anticipation, brey2012anticipatory, clarke2016anticipation, friedman2010multi, friedman2019value,nathan2008envisioning, fox2017imagining, kozubaev2020expanding, auger2013speculative, hong2021planning, ballard2019judgment, jirotka2017responsible, d2021moral}. For example, value sensitive design is a theoretically-grounded design paradigm that can be applied to develop technologies that uphold the values of all potentially impacted stakeholders by identifying and addressing those values early in the technology development process \cite{friedman2013value, friedman2010multi, friedman2019value}. \textit{Envisioning Cards}, a practical tool derived from value sensitive design, can stimulate reflection and dialogue about the long-term impacts of new technology on different stakeholders \cite{nathan2008envisioning}. Through a series of workshops with industry technology developers, Elsayed-Ali et al. iteratively designed and evaluated \emph{Responsible \& Inclusive Cards} aimed at promoting critical reflection on the potential impacts of industry technology work \cite{elsayed2023responsible}. Other common methods such as design fiction \cite{fox2017imagining, markussen2013poetics, blythe2014research} and speculative design \cite{elsden2017speculative, kozubaev2020expanding, auger2013speculative} are often used by CSCW and design researchers to imagine and create future scenarios and objects, in order to help researchers and practitioners explore, reflect, and anticipate the potential societal impact of technology. For example, \textit{Timelines} is a design activity that assists technology developers in reflecting on the ethical concerns related to technical developments by creating fictional stories from news headlines \cite{wong2021timelines}. More recently, HCI researchers have explored leveraging generative AI in scaffolding computing researchers and technology developers in anticipating the potential harms caused by their work \cite{buccinca2023aha, wang2024farsight, pang2024blip}.  With few exceptions (such as the work of Pang et al. \cite{pang2024blip}, which targets researchers within the academy), these tools are primarily targeted at practitioners, rather than researchers. Our work draws inspiration from these works to better understand the unique challenges faces by \textit{industry computing researchers} specifically and to develop empirically-informed recommendations and tools for this distinct community. \looseness=-1

Recently, impact assessments have emerged as a particularly common approach to encouraging proactive anticipation and potential mitigation of harms in computing innovations \cite{moss2021assembling}. These emerging impact assessments for computing innovations often draw inspiration from impact assessments that have already been deployed to attempt to capture various aspects of responsible technology creation, such as fiscal impact~\cite{kotval2006fiscal}, environmental impact~\cite{morgan2012environmental, cashmore2004interminable}, data protection~\cite{PIA_GDPR}, and privacy~\cite{PIA_Canada}. For example, the Chief Information Officers Council of the U.S. and the Government of Canada have each developed a questionnaire-based algorithmic impact assessment as a resource to guide organizations and institutions in ``assess[ing] and mitigat[ing] the impacts associated with deploying an automated decision system'' \cite{AIA_Canada, AIA_CIO}. In collaboration with the Ada Lovelace Institute, the UK National Health Service AI lab piloted an algorithmic impact assessment to help developers ``maximise the benefits and mitigate the harms of AI technologies in healthcare'' \cite{AIA_NHS}. Similarly, in the private sector, Microsoft recently began requiring product teams to complete a responsible AI impact assessment prior to the launch of any AI-based features \cite{RAIIAguide_MSFT, RAIIAtemplate_MSFT}. 

Existing impact assessments for computing innovations are designed for technology developers, often for deployed technologies that have clear use cases in mind \cite{moss2021assembling, bogucka2024co}. These tools are not immediately appropriate for assessing the impact of research publications or other research artifacts, which may be harder to predict since the potential use cases are more open-ended. Furthermore, there is limited empirical knowledge on designing usable and effective tools for computing researchers to assess societal impacts. \cite{do2023s}. In this study, we aim to bridge this gap by iteratively designing and developing an \textit{empirically-informed} impact assessment template for industry computing researchers, drawing from existing tools and grounded in empirical evidences from our co-design sessions. \looseness=-1

\subsection{Understanding and Supporting Responsible Computing in Industry Settings} \label{Industry RAI}

Within the CSCW and broader HCI communities, there has been a significant push to enhance our understanding of industry practitioners' current practices, challenges, and needs around responsible technology design, and to develop tools and guidelines to support them~\cite{rakova2021responsible,passi2018trust,madaio2020co,heger2022understanding, holstein2019co, wang2023designing, deng2022exploring, widder2023s, passi2019problem, deng2023investigating, wong2021tactics, yildirim2023investigating, piorkowski2021ai, zhang2020data, shneiderman2020bridging, sadek2024guidelines, sherman2024ai}. This research has underscored how organizational dynamics and culture influence responsible and ethical computing practices. For example, a culture that predominantly values moving fast and scaling up can obstruct the use of tools designed to encourage thoughtful reflection~\cite{madaio2020co, madaio2022assessing, deng2023investigating, widder2023dislocated}. Moreover, power dynamics in the workplace can significantly affect responsible innovations and technology developments. For example, recent works by Rakova et al.~\cite{rakova2021responsible} and Wong et al.~\cite{wong2021tactics} each revealed how ``ethics workers'' within companies frequently encountered pushback from leadership when advocating for more responsible technologies. Researchers have also developed tools and processes to support workers' reflexive and anticipation work in industry practice, including the Responsible \& Inclusive Cards mentioned above \cite{wang2024farsight, elsayed2023responsible, ballard2019judgment}. On the practical side,  works have further emphasized the importance of integrating support for responsible development into teams' existing tools and workflows~\cite{madaio2020co,heger2022understanding, deng2022exploring, wang2023designing, zhang2020data, wang2024farsight, liang2024s}. \looseness=-1

However, the vast majority of this prior research has focused on \textbf{teams creating and deploying products} as opposed to \textbf{industry researchers} producing academic publications or other research artifacts like data or code. In our study, we specifically investigate how organizational dynamics and company culture might influence industry researchers' capacity and willingness to assess, articulate, and address the potential negative societal impacts of their work. Through interviews and co-design with industry researchers, we develop an impact assessment template tailored specifically for industry research rather than product design and development.

\section{Methods}
\label{sec:methods}

To investigate our research questions, we conducted a series of interviews with 25 industry computing researchers, an iterative co-design process with the same 25 industry computing researchers and 16 impact assessment experts, and a user study with 15 interns working in industry research. We illustrate our method in Figure~\ref{fig:method}. Each activity is described below, and full protocols are provided in the supplementary material.\footnote{The supplementary material will be available with the full published version of this paper.}

All studies were conducted over a video conferencing platform between June and August 2023. Sessions were recorded except in cases in which participants opted out and were transcribed using transcription software.  Participant quotes were anonymized at the individual, team, and organization levels.  Participation in the studies was voluntary and participants were told that they were free to skip any questions they were uncomfortable answering and to leave the study at any time for any reason. Participants in each study were compensated with a \$50 gift card. All studies were approved by our institution's IRB.

 \begin{figure*}[t]
  \centering
  \includegraphics[width=1\linewidth]{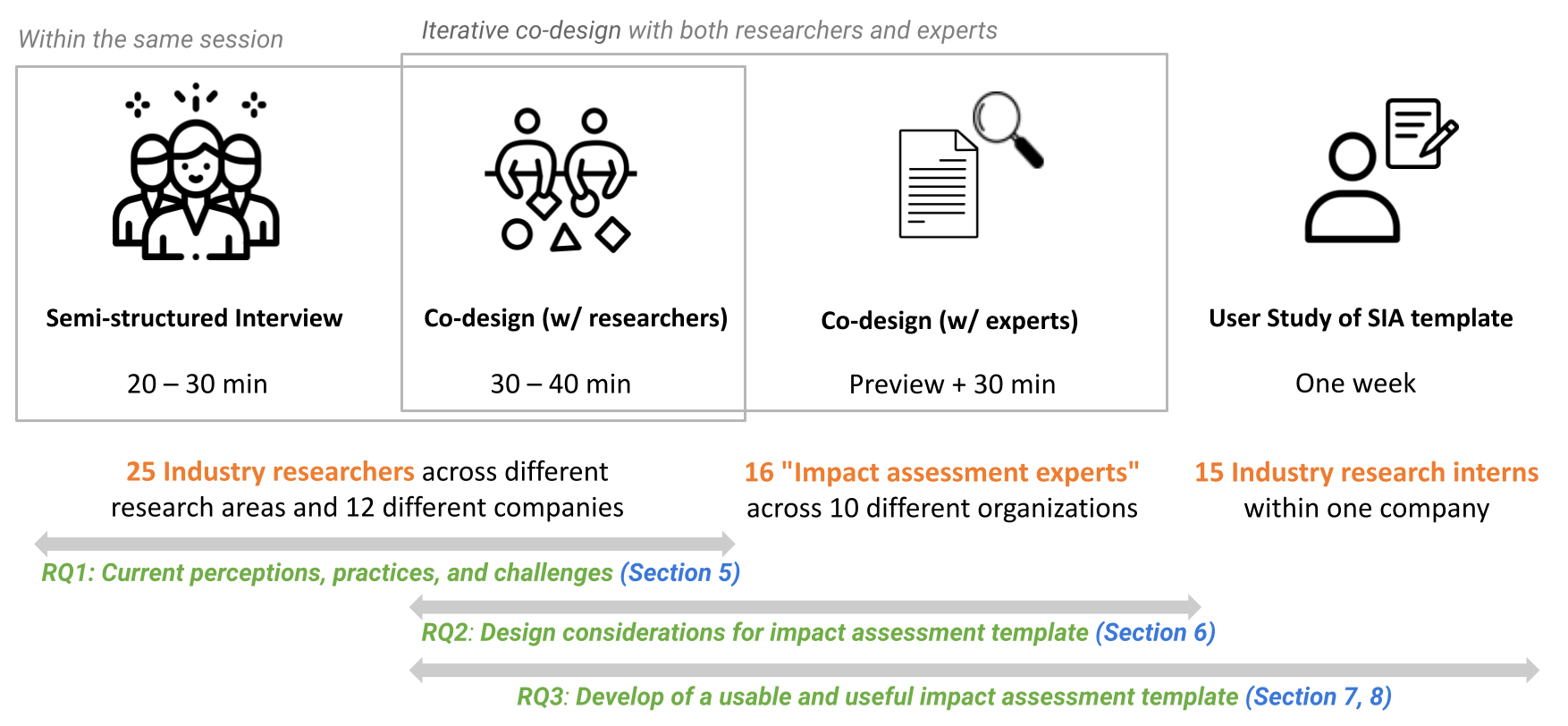}
  \caption{
   Overview of our methods and how each study component maps to our research questions and results.
}
  \Description{TBA }
  \label{fig:method}
\end{figure*}

\subsection{Creating the Initial Prototype of the Impact Assessment Template} \label{initial template}

To create the initial prototype of the societal impact assessment template that we used as a basis for iteration in the co-design sessions, we began by synthesizing insights from prior academic literature and other reports on assessing the impact of research in areas such as AI, HCI, and medical research \cite{andrade2021ai, ada2022looking, AIA_Adalove, metcalf2021algorithmic, olteanu2023responsible}. From these, we formulated a list of initial design considerations. We also drew inspiration from publicly available impact assessment templates such as the responsible AI impact assessment template from Microsoft \cite{RAIIAguide_MSFT, RAIIAtemplate_MSFT}, the algorithmic impact assessments from the UK NHS AI lab \cite{AIA_Adalove}, the CIO Council of the U.S. \cite{AIA_CIO}, and the Government of Canada \cite{AIA_Canada}, and the privacy impact assessments from the Government of Canada \cite{PIA_Canada} and introduced by the EU with the General Data Protection Regulation (GDPR) \cite{PIA_GDPR}. When designing the template scaffolding content, we also drew inspiration from artifacts in value sensitive design \cite{TarotCards, nathan2008envisioning, wong2021timelines, zhu2018value, friedman2019value, shen2021value}, tools and guidelines for conducting responsible computing research \cite{smith2022real, ACL2023ethicspolicy, Neurips2021checklist, hong2021planning, ICML2023EthicsGuidelines}, as well as survey papers discussing the taxonomy of socio-technical harms \cite{shelby2023sociotechnical, Weidinger2022TaxonomyOR}. The resulting template, included in the supplemental material, was intended to guide industry researchers through a step-by-step process to consider the intended uses and applications of their work, identify potential stakeholders, anticipate positive societal impact, anticipate potential negative societal impact, and formulate possible mitigations.

\subsection{Interview and Co-Design Sessions with Industry Researchers} \label{interview+codesign}

\begin{table}
    \centering
    \begin{tabular}{|c|c c||c|c c|} \hline  
         \textbf{ID}&  \textbf{Years} &  \textbf{Research Area}&  \textbf{ID}&  \textbf{Years}& \textbf{Research Area}\\ \hline  
         P01&  0--3&  Augmented Reality (AR)&  P14&  3--10 & VR, AR\\ \hline  
         P02&  3--10 &  HCI, Responsible AI (RAI)&  P15&  0--3& Computer Vision\\ \hline  
         P03&  0--3&  Reinforcement Learning&  P16&  3--10& VR, AR\\ \hline  
         P04&  3--10 &  Accessibility&  P17&  10+& Reinforcement Learning\\ \hline  
         P05&  10+ &  Hardware security&  P18&  3--10 & HCI, Responsible AI\\ \hline  
         P06&  0--3&  AI, Philosophy&  P19&  3--10 & Sociology, CSCW\\ \hline  
         P07&  0--3&  Virtual Reality (VR), Accessibility&  P20&  0--3& NLP, RAI\\ \hline  
         P08&  10+&  AI, Accessibility&  P21&  3--10& AI, HCI, Healthcare\\ \hline  
         P09&  3--10 &  Computational Biology&  P22&  3--10& NLP, Computer Vision\\ \hline  
 P10& 3--10 & Psychology, CSCW& P23& 0--3&HCI, NLP\\ \hline  
 P11& 0--3& Computer Vision, Accessibility& P24& 0--3&Computer Graphics\\ \hline  
 P12& 3--10& Natural Language Processing (NLP)& P25& 10+&Computational Social Science\\ \hline  
 P13& 3--10 & NLP, Education& & &\\ \hline 
    \end{tabular}
    \caption{Details about the industry researchers who participated in our interview and co-design study. Overall, the researchers work at 12 companies: 10 large technology companies with 25,000 or more employees and two smaller technology companies with 5,000--24,000 employees.} 19 researchers are from the U.S., 4 from Europe, and 1 from Australia.
    \label{tab:researchers}
\end{table}

After creating the initial prototype of the template, we began conducting sessions with 25 industry researchers across organizations and research areas. Each 60-minute session was broken into two components: a semi-structured interview and a co-design activity. Ideally, we would have conducted all interviews first and then invited back the same group of participants for co-design sessions. However, we decided to conduct interviews and co-design sessions together due to the challenges of participants declining to return for a separate co-design session (as documented by prior work \cite{lee2021landscape, deng2023investigating, rakova2021responsible}) and limited time for conducting the study.
During the early stages of the project, we placed more emphasis on the semi-structured interviews. Later, we shifted more of the session time to the critique and co-design of the template itself. Between sessions, we updated the impact assessment template based on feedback from participants, as shown in Figure~\ref{fig:iterations} and detailed in the appendix.
This iterative co-design process drew parallels with methodologies deployed in prior work \cite[e.g.][]{madaio2020co,holstein2019co,smith2022real, deng2023understanding, elsayed2023responsible}. \looseness=-1

The \textbf{semi-structured interview} was designed to explore industry researchers' perceptions, current practices, and challenges around assessing potential negative societal impact of their work. We asked participants about their research, their research team, and their relationships with product teams. We then asked them to define the societal impact of industry computing research. Adopting a directed storytelling interview approach similar to prior work \cite{evenson2006directed, davidoff2006principles}, we guided participants to reflect on their current practices around whether, when, and how they (try to) assess and mitigate the potential negative societal impact of their work. Finally, we probed deeper into the practical challenges they had encountered. \looseness=-1

The \textbf{co-design sessions with researchers} were intended to better understand the opportunities for supporting researchers in conducting impact assessment. Within the session, participants were asked to bring a publicly available (e.g., previously published or available on a public platform like arXiv) research paper they had authored as a case study to keep in mind when walking through the template. We first gave participants a high-level overview of the impact assessment template. We then went through the template, asking participants to consider how they would answer the questions with respect to their projects and provide item-level feedback, for instance, identifying sections of the template that might be hard for their team to use.  For each component of the template, we also encouraged participants to share what they thought might be helpful for them to better complete the task. We asked participants to share at which research stages would they want to use the template in their own research, as well as how they envision the outcomes of the impact assessment being presented to other researchers, the product teams in their organizations, and the public. Note that we made a conscious decision to focus on designing the content of the impact assessment, deliberately choosing not to concern ourselves with how this would be implemented in an interactive tool. This decision was primarily based on the understanding that the format of interaction would likely need to be tailored specifically to the organization and its context of use. We address this topic as future work in Section \ref{dis: beyond SIA}. \looseness=-1

To recruit industry researcher participants, we adopted a purposive sampling approach~\cite{campbell2020purposive}. Our aim was to recruit computing researchers who (1) are currently employed in industry and (2) publish and present their research findings and innovations in peer-reviewed venues such as academic conferences or journals.  We allowed two exceptions to the first criterion, allowing two participants (P21 and P25) who currently work in academia but spent extensive periods of time in industry research labs before becoming professors. We believe their experiences could contribute insight into the challenges specific to industry settings.
Participants were recruited through direct contacts at large technology companies, email lists within industry research labs, and snowball sampling. The 25 researchers who participated spanned 12 technology companies, work in diverse areas of computing research, and possess various levels of experience. Table~\ref{tab:researchers} provides an overview of the participants' research area and relevant experience.

 \begin{figure*}[t]
  \centering
  \includegraphics[width=1\linewidth]{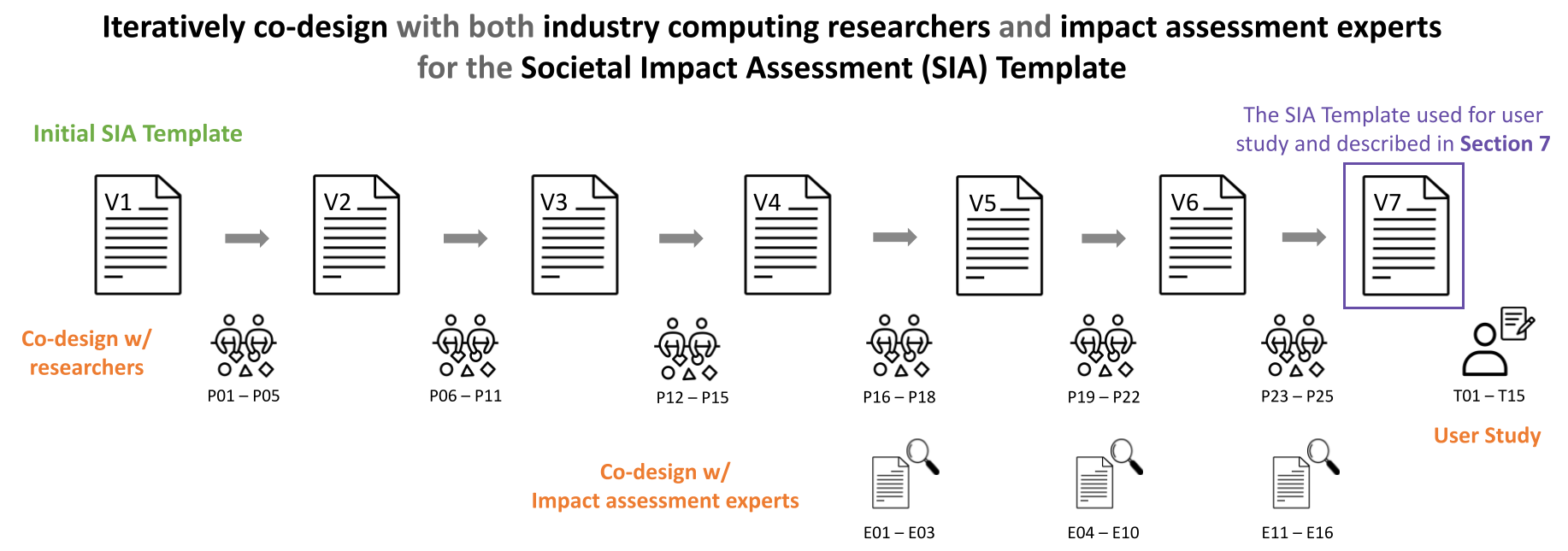}
  \caption{Overview of the iterative co-design process. Version 7 is the SIA Template  described in Section \ref{SIA template} and used in our user study. Version 1 and Version 7 are included in the supplementary materials.}
  \Description{TBA}
  \label{fig:iterations}
\end{figure*}

\subsection{Co-Design Sessions with Impact Assessment Experts}

To complement the insights from industry computing researchers --- the target users of impact assessment templates --- we also conducted a series of 30-minute \textbf{co-design sessions with impact assessment experts}: researchers and practitioners with prior experience in developing and reviewing impact assessments or research or policy-making experience around responsible computing and technology more broadly
(See Table \ref{tab:experts}).
The goal of these sessions was to ensure that the impact assessment template fulfilled its purpose in uncovering potential negative societal impacts. We began conducting these sessions after making several iterations on the template and interleaved them with the co-design sessions with industry researchers, as illustrated in Figure~\ref{fig:iterations}. In preparation for each expert co-design session, we shared a current version of the template with the participant and gave them the opportunity to optionally review and annotate it with any preliminary feedback. During the session, we first solicited feedback on the template's structure and goals. We then prompted the participant to critique each section and provide fine-grained feedback. Finally, we asked the participant to envision how the template might be operationalized in their current organization. We incorporated participants' feedback when producing the next iteration of the template. \looseness=-1

To recruit participants, we again adopted a purposive sampling approach \cite{campbell2020purposive}. In particular, the authors leveraged their connections with impact assessment experts across academia, industry, and civil society. Among the 21 impact assessment experts we emailed, 16 expressed interest. Of these, 14 participated in a full 30-minute session, while two opted to review the template and annotate it with their feedback without joining a live session. Overall, five of the experts were from civil society organizations, five from academia, and six were employed in industry. 
Table \ref{tab:experts} offers more background information on these experts.

\begin{table}
    \centering
    \begin{tabular}{|c||c|c|c|} \hline 
         \textbf{ID}&  \textbf{Organization}&  \textbf{Research/Practice Domain}& \textbf{Relevant Experience}\\ \hline 
         E01&  Industry&  Responsible AI (RAI); Healthcare& Internal impact assessments\\ \hline  
         E02&  Civil society&  Public policy; RAI& Internal impact assessments \\ \hline 
         E03&  Academy&  Science and Technology Studies; RAI& Other impact assessments\\ \hline 
         E04&  Civil society&  AI safety and governance& Internal impact assessments\\ \hline  
         E05&  Academy&  Philosophy of science& Other responsible computing\\ \hline 
         E06&  Industry&  Public policy; RAI& Internal impact assessments\\ \hline
         E07&  Industry &  RAI; Healthcare& Internal impact assessments\\ \hline 
         E08&  Civil society&  Privacy; RAI& Other responsible computing\\ \hline 
         E09&  Civil society&  Public Policy; RAI& Other impact assessments\\ \hline 
 E10& Academy& Philosophy& Other responsible computing\\ \hline 
 E11& Industry& Public policy; RAI&Internal impact assessments\\ \hline 
 E12& Industry& Public policy; Philosophy&Internal impact assessments\\ \hline 
 E13& Academy& Bioethics; Political science&Internal impact assessments\\ \hline 
 E14& Academy& Philosophy; RAI&Other responsible computing\\ \hline 
 E15& Industry& AI safety and governance&Internal impact assessments\\ \hline 
 E16& Civil society& AI safety and governance& Other impact assessments\\ \hline 
    \end{tabular}
    \caption{Details about the impact assessment experts who participated in our co-design study. All have experience developing and reviewing impact assessments in their own organization (denoted ``Internal impact assessments''), working with impact assessments in other contexts (``Other impact assessments''), or in research or policy-making on responsible computing more broadly (``Other responsible computing''). }
    \label{tab:experts}
\end{table}

\subsection{User Study of the SIA template} \label{user study}

By the time we produced V7 of the template, we had reached data saturation with the co-design activities with researchers and impact assessment experts. Similar to previous co-design studies \cite{madaio2020co, deng2023understanding, holstein2019co}, we stopped iterating on the template at this point and turned to a \textbf{user study}. While we do not view V7 as a ``perfect'' or ``final'' version of the template, the user study was intended to give us more insight into the template's usability and usefulness in practice.

Participants in the user study were given one week to complete V7 of the template with respect to an ongoing research project. After completing the template, they were given an exit survey and asked to sign up for an exit interview. The exit survey first asked questions about their usage of the template, such as the amount of time they spent filling it out and how many times they revisited it during the one-week study period. It then asked about the template's overall usability, for example, asking participants to provide Likert scale ratings for statements like ``the template's organization is logical and intuitive to navigate.'' Next, participants were asked whether the template helped them identify potential stakeholders, positive societal impacts, and negative societal impacts. Finally, they were asked for Likert scale ratings of the template's overall usefulness. In the exit interview, which lasted 15--40 min, participants were asked to share any additional thoughts on their experience filling out the template and probed more deeply about their survey responses.

For the user study, we recruited current research interns (primarily PhD students) from a large U.S.-based technology company.  Ideally we would have recruited full-time industry researchers across organizations satisfying the same criteria as the participants in the interview and co-design sessions. However, some researchers were reluctant to join because of the potential risk of getting scooped by sharing their ongoing, unpublished research projects. Additionally, full-time researchers were hesitant to make the required time commitment over the course of a week. We believe that recruiting research interns still allowed us to gain valuable insights on the usability and usefulness of the template. We discuss these limitations further in Section~\ref{limitations}. 

To recruit the participants, we shared a recruitment message in email lists and message boards within the technology company. Initially, 17 research interns signed up for the study, of which 15 filled out the template, completed the exit survey, and participated in an exit interview. Table \ref{tab:interns} provides more information about the participants in the user study.

\subsection{Data Analysis} \label{data analysis}

In total, the interviews, co-design sessions, and user study exit interviews yielded approximately 29 hours of recorded audio, which was automatically transcribed, as well as detailed notes taken by the interviewer for the 11 sessions in which the participants opted out of recording. We additionally had the two annotated templates from the impact assessment experts who chose to provide asynchronous feedback and responses to the 15 exit surveys from the user study. To analyze the transcripts from our interviews and co-design sessions, we adopted the reflexive thematic analysis approach described by Braun \cite{braun2019reflecting}, similar to prior work examining data from interview and co-design activities \cite{madaio2020co, holstein2019co, elsayed2023responsible}. All of the authors convened frequently throughout the interviews and co-design sessions to conduct interpretation sessions and discuss the insights. The first author conducted open coding of the transcripts after each interpretation session. 
Some open codes that emerged from the interviews and co-design sessions with industry computing researchers include: "Participants deflect the responsibility of assessing and addressing the impact of their research to product teams" and "Participants tend to oscillate between identifying stakeholders and assessing the potential impact when co-designing the template." Following the guidelines provided by Braun and Clarke \cite{braun2006using}, during this coding process, authors continuously discussed discrepancies in interpretation, and the first author iteratively refined the codes based on these discussions \cite{braun2019reflecting,mcdonald2019reliability}. For the user study, we analyzed the exit survey data and computed descriptive statistics. We also performed open coding of the open-text responses to the exit survey and the transcripts of the exit interviews, aiming to synthesize qualitative insights that complement the quantitative evaluation provided by participants. Notice that we chose \textit{not} to perform and report a content-wise analysis of the template like prior work \cite[cf.][]{nanayakkara2021unpacking, liu2022examining} because doing so would reveal details about participants' ongoing, unpublished research projects. \looseness=-1

\section{Current perceptions, practices, and practical challenges} \label{empirical findings}

In this section, we identify and explore four factors that influence how industry researchers grapple with (or choose not to grapple with) the potential negative societal impact of their research, summarized in the left part of Figure \ref{fig:results}. Some of our results validate findings from prior empirical work~\cite{do2023s, ada2022looking, bernstein2021ethics, liu2022examining, nanayakkara2021unpacking}; we call out these connections to prior work when they arise. contributes the necessary empirical evidence to be confident of the generality of these observations. Our findings also extend the existing literature by highlighting structures and dynamics that are unique to computing research in industry.
\looseness=-1

\subsection{Relationships between Industry Researchers and Product Teams}
\label{relationship with product}

Overall, we found that the complex relationships between industry researchers and product teams heavily shape participants' perceptions and current practices around assessing, articulating, and addressing the negative societal impact of their work. In particular, many participants believe that their research has impact through (and only through) its incorporation into products and services. This belief motivates them to attempt to grapple with its potential negative societal impact. However, participants often encountered practical challenges in doing so due to a lack of both transparency and agency around the specifics of when and how their research might be used by product teams.  These practical challenges eventually led to participants deflecting the responsibility of grappling with the potential negative societal impact of their own research to product teams and other decision makers within their organizations. Below, we describe these results in more detail.

Many participants expressed a belief that \textbf{societal impact arises when (and only when) industry research is incorporated into products and services}. P04, P07, P11, P15, P18, and P23 all explicitly mentioned that their own research would have only limited impact if it were not integrated into their companies' products and services. P11 told us that their \textit{``research impact is defined by the product impact,''} and that industry research can be more impactful than work done in academia due to the \textit{``potential to improve the quality of life for many people through [the] company's products.''} Similarly, P13, P21, and P25 --- who all had spent multiple years as tenure-track professors in academia in addition to researchers in industry labs --- emphasized impact through products and services as a differentiating factor between academic and industry research. As P21 put it: \emph{``In school, you mostly hear people talk about impact as coming up with creative new ideas, getting accepted to conferences, getting grants, [and] having lots of citations. But in a company, my colleagues and I mostly get really excited when your ideas can actually be in the hands of users through products...that's how the impact is being defined there.''}

To this end, we found that \textbf{participants are particularly motivated to reflect on and assess the potential negative societal impact of their research when they know it will be incorporated into products and services}. For example, P17 shared with us that their team was \textit{``just doing [their] own thing''} when their work \textit{``was largely theoretical,''} but \textit{``started to look into what could go wrong if our research is being widely adopted'' }when the ML systems they built were incorporated into products. Similarly, P11 suggested that in their organization, conversations about the negative impact of their research often \textit{``happened when a research project was being integrated into a product feature...otherwise we somehow feel like this conversation can wait.''} They further acknowledged that prioritizing only research that is being incorporated into products is not a best practice, but their team needed to \textit{``sort out priorities''} given limited time and resources.

In contrast, several \textbf{participants with limited contact with product teams rarely assessed the societal impact of their work and believed it was unnecessary}. For example, P24, who works on improving the computational efficiency of graphics rendering algorithms and 3D modeling, did not see a need to think about the impact of their work on society \textit{``because the research outcomes are tools for other researchers...not used directly by our customers.''} In addition, when the incorporation of research into products and services effectively motivated participants to consider potential negative societal impact, we found that participants \textbf{frequently limited their focus to existing customers and intended stakeholders}, neglecting indirect stakeholders or other segments of society who might be impacted by their research. 

Interestingly, many participants shared that \textbf{industry researchers do not necessarily know specifically when, where, and how their research will be incorporated into the products, leading to practical challenges for assessing and addressing negative societal impact}. For example, P04, who works closely with product teams on improving the accessibility of mobile apps, told us there were often \textit{``non-disclosure agreements among teams''} that essentially left them out on knowing \textit{``whether or not [their] research will eventually be adopted by product teams.''} Because of this, P04 suggested that it wouldn't be feasible for them to assess the potential societal impact of their research. Even those researchers who were certain their research would be used by a product team often reported a lack of agency regarding the way in which it would be incorporated. For example, P21 said: \textit{``Once [the product teams] take [the research results], they can really do whatever...and in marketing, they can spin it however they want...I mean you would hope that they would use it responsibly, but really you basically lose control.''} Similarly, P23, who works closely with a product team implementing a recommendation system, told us that they were \textit{``pretty much out of touch with [their] own work''} once the product team started to incorporate it. This participant suggested that they wish they could \textit{``sit in on some early meetings discussing how our research is being used''} so that it would be feasible for them to \textit{``flag anything that might cause downstream harms.''} 

To this end, we found that many participants tend to \textbf{deflect the responsibility of assessing and addressing the potential negative societal impact of their own research to product teams and other decision makers within their organizations}. For example, P23 suggested that \textit{``it would be unfair to expect us to mitigate harms when the product people are the ones who misuse our research.''} Similarly, P11, who works on machine learning models, believed that product teams should be accountable for assessing any negative impact of their research when it is incorporated into products since they are the ones \textit{``deciding what to do with the models we provide.''} P04, P05, P16, P18, and P21 all explicitly stated their belief that the primary responsibility for assessing and addressing the potential negative impacts of industry research should rest with product teams, again citing researchers' lack of transparency and agency.

\subsection{Organizational Dynamics and Cultures that Prioritize Innovation and Speed} \label{organizational dynamics}

A rich line of prior work in the CSCW and broader HCI community has underscored how organizational dynamics and culture influence industry product teams' responsible computing practices \cite{holstein2019co, madaio2020co, madaio2022assessing, deng2022exploring, deng2023investigating, deng2023understanding, rakova2021responsible, widder2023s, passi2018trust}. In line with this, our study reveals how industry \emph{researchers} are shaped by organizational dynamics and culture --- particularly the prioritization of innovation and speed --- when assessing, articulating, and addressing the potential negative societal impacts of their work. \looseness=-1

To start with, many participants shared that they were \textbf{reluctant or even unwilling to assess and discuss the potential negative societal impact of their work because of an organizational culture of emphasizing the positive side of innovation}. For example, P02 reflected that the \textit{``predominant narrative focuses on how many great things the company will contribute to the world.''} Because of this, they believed that industry researchers often inadvertently viewed their work through \textit{``rose-tinted glasses''} and felt \textit{``peer pressure to always signal how [their] own research agenda is somewhat aligned with [the] company's mission.''}  In some cases, researchers feared retribution for speaking out about negative impact. For instance, P16 commented on how previous layoffs of industry researchers working on AI ethics had a chilling affect on them and their colleagues, dissuading them from sharing potential negative societal impacts with their managers. 

Among all participants, those in \textbf{junior roles} (P01, P15, P23) \textbf{expressed particularly strong reluctance to assess and discuss the potential negative societal impact of their work}. For instance, P15 told us that they felt pressure to \textit{``establish myself and show that my research is valuable for the company by improving the product.''} Although they considered the potential misuse of a face-swapping computer vision algorithm they were working on, especially by users under 18, this participant ultimately did not voice these concerns, stating \textit{``I don't want to be a buzz killer when my manager is very excited about something.''} Similarly, P23 shared with us that \textit{``I'm still trying to figure out my role in the team. At least at this stage, what I talk most about to people is how my research can help them instead of the harms my work can do.''} Although this participant recognized the importance of assessing potential negative societal impacts, they felt it might conflict with their \textit{``career development as the most recent employee of the team.''}

Furthermore,\textbf{ the \textit{``move fast and ship products''} culture} \cite{madaio2022assessing} \textbf{in technology companies also introduces practical challenges for researchers to appropriately assess and address potential negative impact}. For instance, many participants noted that some computing researchers in their companies tend to \textit{``move fast and do groundbreaking innovation'' } (P16), while some product teams aim to \textit{``launch new product features as fast as possible''} (P20). P13 recounted an instance where their research team developed a computer vision algorithm that \textit{``took a very binary gender view of the world''} and they \textit{``didn't feel comfortable releasing it until [they] could do something about that.''} However, they and their colleagues felt \textit{``internal pressure from the product manager to get things out quickly.''} This participant concluded that \textit{``the pressure to get things out quickly is often at odds with carefully assessing the unintended consequences of it.''} \looseness=-1

As a result of this culture, participants reported that \textbf{their efforts to assess and address negative societal impact were often ignored or pushed back on due to mismatched incentives}. Specifically, many participants, including P04, P07, P08, P10--P13, P19 and P21, shared stories about how their colleagues from other teams (both in research and product development) often overlooked advice from researchers on assessing and mitigating negative societal impacts. P10, for instance, recounted a time when a team integrated facial recognition algorithms into tools developed for internal workers, disregarding the internal guidelines suggested by their research group. The participant described that the development team is \textit{``encouraged to move forward without any ethical guidance...they're not incentivized in any way, shape, or form to be ethical. They are just trying to create a prototype as fast as possible, and because they report to a general manager who is not in research, they can basically ignore us.''} Similarly, P08 shared that by the time they published a paper and highlighted the limitations of their AI models, which were designed for an alt-text generation pipeline, the product team working with them had already incorporated their research into new features to achieve the team's quarterly goal. Reflecting on this, P8 stated \textit{`` we spend time writing a paper and wait for peer review, but [product teams] won't wait for our publications...they generally don't read our paper or guidelines. We need ways to affect their day-to-day practice, and the way that you affect their day-to-day practice is through these various compliance programs, or even their KPIs.''} \looseness=-1

\subsection{Disagreements and Misconceptions around Societal Impact} \label{misconceptions}

A number of our study participants believed that \textbf{assessing the negative societal impact of their work was unnecessary due to their specific areas of research}. Echoing prior work~\cite{do2023s, ada2022looking,liu2022examining}, we heard this from participants whose research contributions were in theoretical areas of computer science (e.g., machine learning theory), as well as those whose primary research goals were aimed at doing good, for example, mitigating the potential harms of other computing research. Other participants felt that assessing the potential negative societal impacts of their research was unnecessary \textbf{due to the highly regulated nature of their research application areas}, such as medicine, finance, and healthcare. P09, for instance, noted \textit{``the only [potential negative societal impact] I could think about is people making bio-weapons...the problem is that building bio-weapons is hard and requires a lot of resources.  It’s a bit unnecessary for me and people to worry about the malicious usage because the government will definitely notice if someone is making my research into bio-weapons.''} P14, who works on VR/AR research in a Fintech company, expressed that there is no need for researchers to complete impact assessments since the stringent regulations in finance would enforce that care would be taken downstream in deployment. In particular, this participant mentioned that since their company has \textit{``security engineering teams to prevent the malicious actors and law professionals to handle any legal disputes or litigation,''} their responsibility as a research scientist is simply to innovate. We note that both P09 and P14 called out only examples of malicious misuse when contemplating potential negative societal impacts. However, during the co-design process, both P09 and P14 identified potential accidental misuses of their research and agreed that there was still value in assessing and articulating these potential misuses of their work. 

All but two participants in our study \textbf{conflated ensuring research integrity and attending to the welfare of human subjects with addressing the downstream societal impact of research}. While prior work has surfaced similar misconceptions~\cite{do2023s, ashurst2022disentangling}, we were particularly surprised by how prevalent such misconceptions were among our participants, with participants of varying seniority levels, working in areas including sociology, human-computer interaction, and responsible AI, all hitting this pitfall. For example, when asked to describe an instance when their team attempted to assess the potential negative societal impact of their research, P10 recounted how their team had protected the privacy of participants outside the U.S. while conducting research. After we clarified the question and the participant realized their misunderstanding, they began to reflect on how their team had also been conflating these concepts during their research meetings: \textit{``I guess we have been kind of mixing these two concepts together in our project meetings...but now that I think about it carefully, protecting participants' privacy is actually different from preventing downstream harms.''} Consequently, P10 believed that it would be beneficial to provide researchers with learning material and training opportunities around disentangling these concepts and appropriately addressing ethical issues throughout the research lifecycle. \looseness=-1

\subsection{Lack of Sufficient Infrastructure } \label{lack infrastructure}
Prior research has already highlighted the severe absence of efficient and effective infrastructure to support computing researchers in assessing and addressing potential negative impacts  \cite{do2023s, ada2022looking, bernstein2021ethics, nanayakkara2021unpacking, liu2022examining, ashurst2022ai, boyarskaya2020overcoming}. Many industry researchers in our study noted that, \textbf{while there are processes within their organizations for brainstorming the \textit{positive} impact of their research, there are often no such processes for assessing and addressing the potential negative societal impact}. For example, P01 told us that their team usually schedules multiple \textit{``research brainstorming sessions,''} often resulting in formal documentation around their projects' potential positive impacts, managed by the team lead. In stark contrast, when assessing their research’s potential negative impact, this participant's team relies on \textit{``spontaneous conversations about what might go wrong in one meeting...maybe some follow-up if someone happens to bring this up again in a later meeting.''} P21 said that while discussions about negative impact occur, there is no formal process, so discussions are often being relegated to \textit{``side chatter.''}

In contrast, some participants mentioned that in their companies, formal processes are beginning to emerge to involve industry researchers in assessing and addressing the potential negative societal impact of their research. These processes often come up in the context of discussions around legal and compliance considerations that occur when the company is releasing research artifacts through open source or incorporating research into products. Many participants pointed out a \textbf{lack of support around cross-functional communication and collaboration between researchers and legal or compliance teams}. For example, P22 shared the communication gaps they experienced when discussing the release of an image captioning model with their in-house legal teams: \textit{``I found it useful when legal folks came to talk to us, but the discussion between us was not efficient...I was very confused about all the terminologies they threw at us...[it] felt like we needed a translator despite both speaking English.''} Similarly, P14 described feeling \textit{``unprepared and confused''} in discussions about the potential misuse of augmented reality services with the company's compliance team, adding \textit{``we all agreed that understanding the platform policies was important to consider the different downstream use cases of our work...but none of us on the research team was aware of those policies before that chat...I wish we talked to them earlier on.''} In light of this, this participant emphasized the need for processes that foster better integrated communication and collaboration between research and compliance teams, particularly for assessing and addressing potential negative impacts. \looseness=-1

Participants also mentioned the lack of support and effective infrastructure from publication venues like conferences that require a discussion of societal impact or an explicit broader impact statement. In line with the hypothesis from prior work that authors often write impact statements without sufficient reflexivity ~\cite{liu2022examining, nanayakkara2021unpacking}, we repeatedly heard from many participants (P01, P02, P05-- P09, P12--P16, P20, P21, P23, P25) that \textbf{they only start considering the broader impact statement close to the paper deadline}. Participants like P13 and P21 expressed a desire for conferences or their own organizations to provide more concrete guidance on assessing and reporting the societal impact of their work. Corroborating findings from prior work \cite{do2023s}, many participants in our study mentioned that, \textbf{without sufficient scaffolding, the conference requirements do not motivate them to conduct a careful impact assessment and can feel burdensome}. For example, P03 told us that a rigid broader impact statement requirement did not fundamentally motivate them to assess the impact of their research, as \textit{``anything that dictates how to structure your paper or mandates content...feels like it's taking away some of your agency.''} P17 suggested that these conference requirements are treated as mere checklists by industry researchers without \textit{``a norm and culture among the community to treat evaluating the downstream impact of our work as the first class problem.''} In addition, P19 noted that adding a conference requirement without providing sufficient training and support inadvertently adds burden to researchers.

\section{Design considerations for societal impact assessment} \label{design considerations}

In this section, we synthesize a list of design considerations (DCs) for the design and implementation of societal impact assessments for computing researchers, as overviewed in Figure \ref{fig:results}. We support each DC with empirical evidence from the co-design process with 25 industry computing researchers (whom we refer to as ``researchers'' for brevity throughout this section) and 16 impact assessment experts (whom we refer to simply as ``experts''). Additionally, we briefly discuss how each DC might be realized in practice, as suggested by both researchers and experts. We present these DCs in two subsections: those that pertain to template content and structure  (Section~\ref{DC for template}), and those that pertain to use within an organization (Section~\ref{DC for organization}). In Section  \ref{SIA template}, we demonstrate how to operationalize these DCs within an actual impact assessment template.  \looseness=-1

\subsection{Design Considerations for Template Content and Structure} \label{DC for template}

\noindent\textbf{DC-1. Provide scaffolding for researchers to consider different dimensions of societal impact.} This scaffolding could take the form of brainstorming examples, checklists, and guided questions for each dimension. \looseness=-1 

In response to the lack of infrastructure for assessing negative societal impact (Section \ref{lack infrastructure}), researchers unanimously expressed a desire for processes to help them grapple with the diversity and complexity of potential impacts. These include the various types of stakeholders possibly impacted, the ways in which the research is used, and the different types of societal impact. P13 mentioned constantly being \textit{``afraid of missing some important elements that need consideration.''} P19 suggested including scaffolds such as checklists to ensure that researchers address all important considerations. \looseness=-1

This sentiment was echoed by all experts. For instance, E07 shared that they observed that researchers often \textit{``stay in their comfort zone''} and \textit{``keep bringing up commonplaces''} when considering potential downstream harms, rather than \textit{``critically examin[ing] how research could unfold in different ways in the real world.''} E03 believed that in addition to considering stakeholders and types of societal impact, the template should also encourage researchers to \textit{``specifically reflect on the long-term and indirect effects of their research on society.''} Multiple experts also emphasized the need to include a list of concrete examples under each dimension, as they are \textit{``really the best way to get someone to think about a risk or aspect that they hadn’t previously considered''} (E16).

\hfill

\noindent\textbf{DC-2. Support heterogeneous, non-linear thinking processes.} Specifically, allow researchers to complete the sections of the impact assessment template non-sequentially, in the order they prefer. \looseness=-1

While many existing impact assessments in other domains instruct practitioners to proceed in a linear manner \cite{AIA_CIO, AIA_Canada, PIA_GDPR}, we observed that researchers frequently jumped between different sections of the template. For instance, P17 oscillated between identifying stakeholders and thinking through how these stakeholders might misuse or be impacted by their research, commenting: \textit{``I wonder if there is a way to design the template so that it's easier to simultaneously consider the stakeholders and the negative impact.''} P20 went back and forth between considering the limitations of the research and brainstorming the potential negative impact. E06 emphasized that researchers need to consider the \textit{``interactions and trade-offs between the categories of impact,''} suggesting that researchers should \textit{``be reminded to reconsider stakeholders while responding to questions in subsequent sections, to prevent overlooking marginalized communities that were not recognized earlier.''}

While most researchers found the identification of potential stakeholders to be a natural starting point for considering potential impacts, others expressed different preferences. P09 suggested they would prefer to \textit{``start with thinking about the type of impact,''} while P23 found it \textit{``more natural to first figure out how people in general might misuse my work, then dive into more details on specific groups of stakeholders.''}
E07 suggested including \textit{``a questionnaire at the beginning of the template to solicit [a] preference, then adapt the [template] structure based on the response.''}

\hfill

\noindent\textbf{DC-3. Support conducting the impact assessment iteratively over time.} Design the template to be revisited and updated throughout the research process, accommodating changes and new insights as the project evolves.

Many researchers brought up that impact assessment \textit{``shouldn't be a one-shot process''} (P16), and that researchers should instead \textit{``fill out this template when starting the project and revisit it from time to time''} (P21).
Similarly, experts advocated for treating the template as a \textit{``dynamic document, revisited as research progresses''} (E09). E05 highlighted the necessity of continuous impact assessment, noting the pitfalls they observed: \textit{``researchers sometimes try to get all questions down in one pass...which shouldn't be the case since they are supposed to iterate on their answers throughout the process...new problems will always emerge as the research evolves.''}
Both researchers and experts suggested that the template could include version control or progress-tracking features so that researchers could \textit{``have iterated documentation available to reflect on how their thinking [about the societal impact] evolve[d] throughout the research process''} (E01). \looseness=-1

\hfill

\noindent\textbf{DC-4. Highlight important considerations for specific research sub-fields and application domains.} Specifically, augment the template with domain-specific examples, terminology, and considerations so that the template is more relevant and effective for researchers within particular fields. \looseness=-1

Throughout the co-design study, all researchers repeatedly told us that they wished the template would highlight important considerations for their specific research sub-fields and application domains. For example, many researchers working in AI suggested that the template should support AI researchers to \textit{``consider different stages of [the] AI lifecycle, such as problem formulation''} (P06). P16 believed it was important to include \textit{``physical health risks and psychological effects unique to our VR research.''} Several experts echoed these concerns. For instance, both E01 and E05 used computing research in healthcare as an example, suggesting the template should remind researchers working in a healthcare context to consider \textit{``what type of hospitals are you working with to get your data? Is it public or private?''} (E01) and \textit{``differences between private health insurance companies and public health systems''} (E05). Other experts suggested that the environmental impact of model training 
should be highlighted in the societal impact template for researchers working on large language models. \looseness=-1

In practice, experts suggested that template designers could start with a general version of the template, then adapt it to different research domains. One way to achieve this is by including additional modules pertinent to specific research areas. Echoing the recommendations made in prior work examining other forms of impact assessments~\cite{moss2021assembling, bernstein2021ethics}, experts in our study highlighted the need to \textit{``incorporate diverse perspectives from different domains''} (E03) and \textit{``assemble a group of researchers in a specific area to tailor the impact assessment template''} (E06) to be suitable for a specific domain. \looseness=-1

\hfill

\noindent\textbf{DC-5. Highlight the value of completing the template.}
This could include surfacing more positive impacts of the research, improving the rigor of the research, and surfacing future work. \looseness=-1

The challenge of incentivizing researchers to complete impact assessments is well-documented \cite{prunkl2021institutionalizing, ashurst2022ai, ada2022looking, bernstein2021ethics}; see Sections \ref{prior challenges} and \ref{lack infrastructure}. During the co-design process, almost all researchers indicated they would be more likely to complete an impact assessment if it had a clear benefit for their research. As a comparison, P25 shared that they are \textit{``a big fan of pre-registration in fields like experimental psychology and medicine...since I know [completing pre-registration] would make my work better.''}
Many researchers found that they appreciated the activities that helped them identify the positive societal impact and limitations of their work, as they are \textit{``something that can directly benefit paper writing''} (P22). Some, including P02, P04, P09, P13, and P19, also recognized the benefit of identifying negative impacts early to improve their problem formulation and research design; P19 suggested the template designers should emphasize this benefit. \looseness=-1

Experts concurred that the template should emphasize its role in enhancing research quality to mitigate the perception that completing an impact assessment is a punitive and burdensome process. E09 recounted that senior leadership in their organization clarified that \textit{``the purpose of impact assessment tools is not punitive but to help improve research quality...and maximize positive impact while minimizing negative ones.''} Both E02 and E13 emphasized that aligning the template with researchers' interests is crucial for gaining researcher buy-in.

\hfill

\noindent\textbf{DC-6. Suggest appropriate and actionable strategies to address potential negative impact.} The template should also assist researchers with the prioritization of impacts.

Resonating with findings from prior work \cite{do2023s}, both researchers and experts emphasized that the template should extend beyond simply identifying negative impacts to provide actionable steps for developing a harms mitigation plan. For example, P02 told us that they would \textit{``not feel comfortable to finish an impact assessment without knowing concrete next steps for mitigating harms.''} E04 told us that when reviewing impact assessments in their organization, they observed that researchers often acknowledge harms without proposing solutions. To this end, E04 believed that it is crucial for the template to \textit{``prompt people in that general direction to think about appropriate mitigation, instead of just pointing out harms.''} Meanwhile, many experts acknowledged the practical difficulty of fully anticipating the likelihood and magnitude of harm. Nonetheless, they believe the template should at least help researchers begin assessing these factors to \textit{``identify high priority risks and prioritize mitigation''} (E08). In addition, the template should aid in allocating responsibility for addressing potential negative impact. For example, when critiquing the template (V5), E05 noted that the template should \textit{``acknowledge that some mitigation might come from the researchers, but some might be from the company or society as a whole.''} Hence, E05 recommended that the template should \textit{``provide space for researchers to write down: who is responsible for the mitigation? If [researchers] are not responsible, do [they] know whose job it is? [Researchers] can offload responsibilities to someone else, but it is [their] job to clearly indicate who is responsible in the assessment.''}

\hfill

\noindent\textbf{DC-7. Be up front about the limitations and scope of impact assessment.}  Limitations include the impossibility of anticipating every potential negative impact, the impossibility of foreseeing long-term impacts prior to applying the research in practice, difficulties in quantifying societal effects, and the subjective nature of interpreting societal impact.

Experts noted that a single research team \textit{``can't exhaust all the potential use cases [of the research] or predict how people from a different culture or region might be impacted by their work''} (E04), and as such, recommended that template designers frame impact assessment as one component of a broader, continuous dialogue.
Many experts stressed that to be effective, self-reflection through the template should lead to further action. E06 suggested that the \textit{``template should be a starting point for researchers to consider who else they should consult.''} Similarly, E15 highlighted that, even after completing the template, \textit{``researchers should collaborate with product and compliance teams to monitor the usage of the research, as some harms may only become evident later.''} 
Echoing prior research~\cite{ada2022looking, metcalf2021algorithmic},  experts advised that template designers should include disclaimers outlining these and other limitations of impact assessment.

\subsection{Design Considerations for Use Within an Organization} \label{DC for organization}

The following design considerations apply when the impact assessment template will be used within an organization, for example, as part of internal compliance processes.

\hfill

\noindent\textbf{DC-8. Clarify the role of the impact assessment template within the organization.} Specifically, clearly articulate how the completed template will be used and shared within the organization, including any influence it may have on the release or use of research findings. Specify procedures for reviewing completed templates and how the results impact project trajectories.

Experts emphasized the importance of clarity on how \textit{``socially the template will be situated within an organization''} (E09). For example,  the template should \textit{``clearly indicate the audience for the questions and ensure appropriate language is used''} (E05), since researchers may be accustomed to academic styles of writing which would not be suitable for communicating with stakeholders like product teams. Understanding the intended audience can also incentivize higher quality responses. E12 shared that in their organization, \textit{``once we told the researchers that their manager will read [the assessment], some C-suite executives will also see it...now we see a drastic improvement in the quality of the assessments.''} \looseness=-1

Experts suggested several key questions to answer when using the template within an organization: At what stage will researchers fill out the template? Who is responsible for completing and reviewing it? Is the process voluntary or mandatory? What incentives or mechanisms are in place to encourage template completion? How does the template integrate with existing organizational procedures? Responses to these questions would vary across organizations. Organizations could potentially develop companion guidelines that cover these questions, similar to Microsoft's Responsible AI Impact Assessment Guide for product teams~\cite{RAIIAguide_MSFT}.

\hfill

\noindent\textbf{DC-9. Enable knowledge sharing around impact assessments across teams within an organization.} This includes sharing outcomes, examples, experiences, and insights.

Similar to what Do \& Pang et al. found in their study with academic computer science researchers \cite{do2023s}, many industry researchers in our study also expressed an interest in learning from the experiences of others, especially those in related research areas or within the same organization. For instance, P07 was curious about \textit{``how other VR researchers plan to monitor abusive behaviors''} and P15 wanted to know \textit{``what other researchers in my company typically include for this question about the possible stakeholders.''} Experts also underscored the importance of enabling knowledge sharing among teams in research, product, and compliance. E02, for example, suggested having researchers document \textit{``challenges faced while filling out the template for some nice self-reflection...and also [to] provide some valuable insights for other researchers.''}

Some researchers mentioned seeing value in their companies hosting a \textit{``workshop to reflect on effective ways to complete the template''} (P16). Resonating with recommendations from prior work \cite{do2023s, bernstein2021ethics}, many experts proposed collecting case studies and news articles on unintended consequences of computing research, alongside \textit{``exemplar assessments to guide those unfamiliar with the impact assessment process''} (E12). Some experts, however, cautioned against relying on repositories of completed templates. E02 and E07 both expressed concerns that this might inadvertently encourage researchers to copy answers without deeply reflecting on their own research.

\hfill

\noindent\textbf{DC-10. Ensure consistency with other existing tools and guidelines within an organization.}
For example, reference and integrate other existing documents and resources within the template.

In line with the findings of past work on responsible computing in industry~\cite{deng2022exploring, yildirim2023investigating, madaio2020co, heger2022understanding, lee2021landscape}, many experts highlighted the need to tailor the template to align with institutional infrastructure. E07 noted while reviewing the template, \textit{``We have known limitations and [...] misuse in the <internal reviewing tool>. I am wondering if there is a way to connect both [the internal tool and this impact assessment].''} When envisioning using the template within their organization, E11 suggested that \textit{``for misuse, I would also suggest adding <common internal term>, because that is the language we use.''}

Experts emphasized that template designers need to be cognizant of existing resources and processes within the organization, ensuring the template complements rather than conflicts with or duplicates them. For example, some experts suggested that the template should \textit{``leverage resources from existing departments or programs in the company''} (E14) or even \textit{``directly plug into an existing process''} (E12). Related to DC-5, experts recommended the template should highlight the additional value it could provide to researchers while ensuring it is coherent with other organizational infrastructure. \looseness=-1

\section{The Societal Impact Assessment Template } \label{SIA template}

In addition to the previous 10 design considerations, the iterative co-design activity also yielded seven versions of the Societal Impact Assessment template  (Figure \ref{fig:iterations}). In this section, we provide a high-level description of Version 7 (referred to simply as the SIA template for the rest of this paper). The SIA template includes a list of scaffolding elements and 16 questions divided into 4 main sections (Figure \ref{fig:SIA}). We include the full SIA template in the supplementary materials and at \url{https://perma.cc/B396-DMTV}. \looseness=-1

As described in Section \ref{initial template}, the scaffolding elements and questions in Version 1 of the SIA template drew on a rich line of prior tools and research. They were then iteratively refined by incorporating feedback from both industry computing researchers and impact assessment experts throughout the co-design process.\textbf{ }The resulting SIA template, along with the previous 10 high-level design considerations, serve as \textbf{two} \textbf{complementary resources} for future researchers and practitioners to build on, perhaps when developing impact assessment tools and processes for their own organizations. In this section, we demonstrate how (most of) the design considerations in Section \ref{design considerations} were fulfilled in the SIA template. Note that DC-4 is not sufficiently addressed by the current SIA template, nor are DC-8 through DC-10, since these pertain to the template's use within a specific organizational context. We discuss these limitations and future improvements in Sections \ref{user study results} and \ref{dis: beyond SIA}. \looseness=-1

\hfill

\noindent\textbf{Onboarding Section.} Because researchers are not necessarily familiar with the process of conducting impact assessments, may not be well-motivated to do so, or might even misunderstand what constitutes societal impact, we provide an onboarding video for researchers who plan to use the SIA template. In the onboarding video, we first highlight the reasons for using the SIA template to motivate researchers to complete it throughout the research life cycle (\textbf{DC-5}). We clarify that the template is not for addressing risks to human subjects in research, a common misconception (Section \ref{misconceptions}), nor is it a substitute for existing institutional review processes (\textbf{DC-10}). While previewing the template's structure (Figure \ref{fig:SIA}), we highlight that researchers may skip back and forth between the three subsections on potential impact to support non-linear thinking (\textbf{DC-2}). Finally, researchers are instructed to document any challenges faced in answering questions and potentially revisit them, instead of skipping them or providing inadequate responses (\textbf{DC-3} and \textbf{DC-7}). \looseness=-1

\begin{wrapfigure}{l}{0.42\textwidth}
  \centering
  \includegraphics[width=0.41\textwidth]{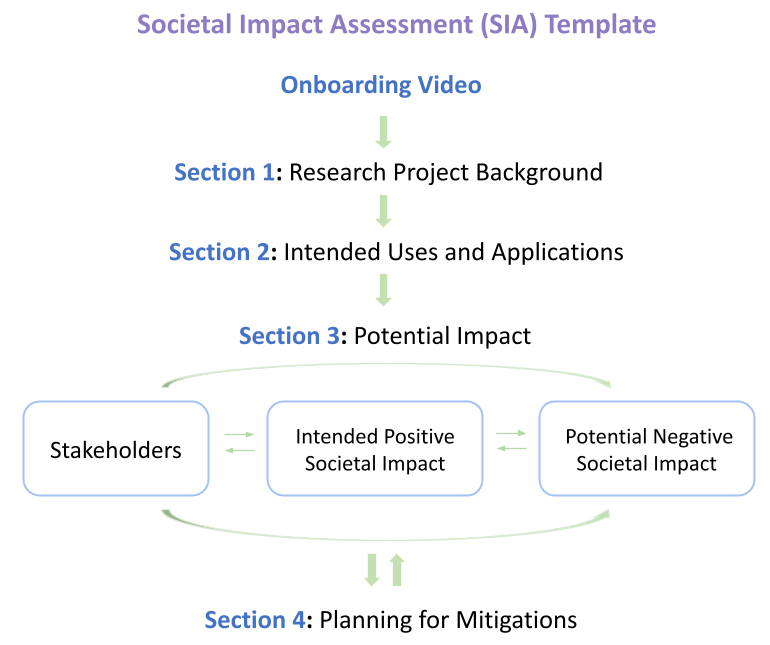}
 \caption{Overview of the SIA template's structure. Please refer to the SIA template included in the supplementary materials for more details on the questions we included in each section and how we guide researchers in thinking through these questions.}
  \Description{TBA }
  \label{fig:SIA}
\end{wrapfigure}

\hfill

\noindent\textbf{Section 1: Research Project Background.} 
To ensure an effective impact assessment, researchers should describe their project in accessible language for the template's intended readers (\textbf{DC-8}).
Therefore, this section first requires a research project description with sufficient details using sufficiently accessible language for readers such as product and compliance teams to understand. We then provide a list of potential assets that researchers might release (e.g., paper, data, code, demos) and prompt researchers to reflect on whether these assets will be shared publicly or only internally within their own institution, as the release of different assets and different release plans can raise different risks (\textbf{DC-6}). \looseness=-1

\hfill

\noindent\textbf{Section 2: Intended Uses and Applications.} 
Because the likelihood of industry research being integrated into products and services might affect the risks raised by the research or the urgency to address them (\textbf{DC-6}), this section guides researchers to share the envisioned applications of the research, with a focus on its relevance to products or services within the organization. This information also serves as a basis for researchers to refer to when considering dimensions of potential impact in the remaining sections. \looseness=-1

\hfill

\noindent\textbf{Section 3: Potential Impact.}
Here we provide a substantial amount of support for researchers to consider various dimensions of societal impact (\textbf{DC-1}). In the template, we highlight that this section should only serve as a starting point for impact assessment, and researchers should revisit their responses as many times as needed throughout their research (\textbf{DC-3} \& \textbf{DC-7}). Finally, we repeatedly remind researchers that they may skip back and forth between the three subsections, as thinking through impacts may require non-linear thinking (e.g., first thinking about possible harms before considering specific stakeholders) (\textbf{DC-2}). As shown in Figure \ref{fig:SIA}, we decided not to number the three subsections to encourage researchers to complete them in any order (\textbf{DC-2}).

\textbf{- Stakeholders.} This subsection scaffolds researchers to identify the possible stakeholders who might use or be impacted by their research, including its future applications. We list individuals (directly and indirectly impacted), marginalized groups, companies, industries, governments, and civil society as potential stakeholders to encourage researchers to think broadly about those who might take up or be impacted by their work. Brief descriptions are provided for each stakeholder category to assist those less experienced in considering these groups.

\textbf{- Intended Positive Societal Impact.} Before diving into  negative societal impact, this subsection prompts researchers to think through the intended positive societal impact of their research, by asking them to describe the best-case scenario for all the impacted stakeholders they envisioned in the Stakeholders subsection. To scaffold brainstorming, the SIA template includes a brief description of each of the following potential types of impact for researchers to consider: economic impact, policy impact, environmental impact, cultural impact, and knowledge impact. \looseness=-1

\begin{wrapfigure}{r}{0.33\textwidth}
  \centering
  \includegraphics[width=0.36\textwidth]{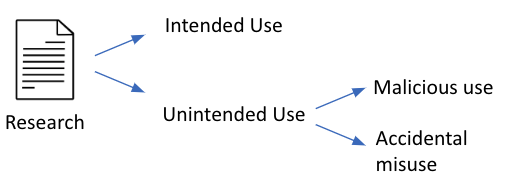}
 \caption{Illustration included in the SIA template to guide researchers thinking through scenarios leading to potential negative societal impact.}
  \Description{TBA}
  \label{fig:research}
\end{wrapfigure}

\textbf{- Potential Negative Societal Impact.} This subsection starts by guiding researchers in considering and documenting the limitations of their research, on the belief that it is helpful to consider how undisclosed or unrecognized limitations can lead to unintentional misuse of research findings and corresponding negative societal impacts.
Researchers are then prompted to contemplate potential harms stemming from misuse (accidental unintended use), abuse (intentional malicious use), and intended use (Figure \ref{fig:research}). We provide concrete examples for each of these three scenarios to support brainstorming. \looseness=-1

\hfill

\noindent\textbf{Section 4: Planning for Mitigation.} This final section guides researchers step-by-step through developing actionable strategies to address potential negative societal impacts surfaced in the previous section (\textbf{DC-6}). The section first helps researchers assess the seriousness of the identified risks by considering the likelihood, magnitude, concentration, and time horizon of the potential negative impacts, and asks researchers to document which risks they view as high and low priority in light of these differences. We then provide a list of methods to help researchers plan possible mitigations throughout different stages of the research life cycle. Finally, we ask researchers how the risks identified as high-priority can be mitigated and what actions they can take to ensure that these mitigations are put in place. To more realistically reflect the scope of what researchers themselves might be able to do to address identified risks (\textbf{DC-7}), we also inquire about steps they can take to encourage others to implement these mitigations if they cannot do so themselves. \looseness=-1

\section{User Study of the Societal Impact Assessment (SIA) Template} \label{user study results}

In this section, we present the findings from the one-week user study, described in Section \ref{user study} . We report the participants' ratings from the exit survey and their comments from their exit interviews, conducted after they had a week to complete the template. For each rating, we report the mean value ($\mu$) and standard deviation ($\sigma$) on a scale of 1 to 5, where 1 represents ``strongly disagree'' and 5 represents ``strongly agree.''

As shown in Table~\ref{tab:interns}, participants were in different stages of their research projects: four in early stages, eight in the middle, and three late stages, writing up results. On average, they spent 81.26 minutes (SD=37.58) filling out the template, with reported times ranging from 25 to 150 minutes. Out of the 15 participants, 11 reported revisiting the template during the one-week user study period: four revisited it once, four revisited it twice, one revisited it three times, one revisited it four times, and one revisited it six times. \looseness=-1

\subsection{Perceived Strengths of the SIA Template} \label{strengths of SIA}

In general, \textbf{participants perceived the template's structure and content as clear and easy to navigate}. 87\% of participants (13/15) agreed (i.e., selected either ``agree'' or ``strongly agree'' in the exit survey) that the template is ``easy to understand'' ($\mu$=4.27, $\sigma$=0.71) and the ``instructions provided with the template are clear'' ($\mu$=4.33, $\sigma$=0.72). A significant 93\% of participants (14/15) agreed that the ``template's organization makes sense and is intuitive to navigate'' ($\mu$=4.40, $\sigma$=0.83). T08 noted in their exit survey that the SIA template presented “\textit{a thoughtfully laid out and thought-provoking assessment. Very relevant to the current state of technological advances in [their] research field.''}

The template also \textbf{effectively facilitated the identification of unconsidered aspects of societal impact, promoting researchers' critical reflection}. Participants rated it highly for the template's utility in identifying potential stakeholders who could benefit from or be impacted by the research ($\mu$=4.27, $\sigma$=0.59 and $\mu$=4.40, $\sigma$=0.74, respectively), as well as potential positive ($\mu$=4.00, $\sigma$=0.93) and negative ($\mu$=4.07, $\sigma$=1.03) societal impacts.  T04 reported that the template prompted them to consider \textit{``minority job candidates who might be affected''} and to rethink how their work should be framed and presented. T13 highlighted in the exit survey that \textit{``it was very helpful to think about stakeholders in an increasingly broader context, starting with individuals and ending at governmental level.''} T07 mentioned in their exit survey that \textit{``I originally thought my research had no negative impact because it was merely assistive tech for blind people...but when I really thought about it when filling it out, I found that there were unintended consequences I never thought of.''} They added during the exit interview that they would update the system design to potentially mitigate these unintended harms. \looseness=-1

Additionally, participants found that \textbf{completing the template was valuable for their research}. In particular, a majority of participants (14/15) felt the template could help them ``draft a societal impact statement to include in my research paper'' ($\mu$=4.40, $\sigma$=0.73). T12 felt \textit{``more prepared to write a blog post about the work after using this template.''} In addition, 73\% of participants (11/15) found ``the template adds value to the research process'' ($\mu$=4.07, SD = 0.79). During the exit interview, many participants, including T01, T03--07, T11, and T13, indicated that completing the template helped them enhance the rigor of their research and identify new research directions.

Finally, completing the template \textbf{improved participants' personal motivation to grapple with the potential negative societal impact of their work.} A notable 80\% of participants (12/15) agreed that “the template convinced (or further convinced) me that I should consider and grapple with the potential negative societal impact of my research” ($\mu$=4.20, $\sigma$=0.86). 
All participants agreed that they would ``recommend this template to others conducting computing research''  ($\mu$=4.27, $\sigma$=0.45). Participants such as T02, T06, T13, and T14 specifically mentioned in both their exit surveys and exit interviews that they would continue using the template themselves for future research projects to assess, articulate, and address potential negative societal impacts. \looseness=-1

\begin{table}
    \centering
    \begin{tabular}{|c||c |c |c|c|l|} \hline 
         \textbf{ID}&  \textbf{Research Area}&  \textbf{Project Stage}& \textbf{Time Spent}& \textbf{Revisit \#} &\textbf{Length}\\ \hline 
         T01&  NLP&  Executing the research&  90 (Minutes) &1 &1,062\\ \hline 
         T02&  Reinforcement Learning &  Executing the research&  25&0 &147\\ \hline 
         T03&  Mathematics&  Early brainstorming&  60&1 &702\\ \hline 
         T04&  Machine Learning&  Executing the research&  75&2 &1,079\\ \hline 
         T05&  AI, Healthcare&  Early brainstorming&  75&0 &1,312\\ \hline 
         T06&  Robotics, AI&  Executing the research&  30&0 &246\\ \hline 
         T07&  Accessibility&  Executing the research&  90&3 &1,189\\ \hline 
         T08&  HCI, Design&  Early brainstorming&  150&2 &702\\ \hline 
         T09&  NLP, HCI&  Executing the research&  60&2 &979\\ \hline
 T10&  Sociology&  Executing the research&  75&1 &672\\ \hline
 T11&  VR, AR&  Preparing publication&  60&0 &1,143\\ \hline
 T12&  NLP, HCI &  Executing the research&  90&1 &418\\ \hline
 T13&  NLP&  Preparing publication&  150&6 &1,136\\ \hline
 T14&  NLP&  Early brainstorming&  90&1 &461\\ \hline
 T15&  NLP, Accessibility&  Preparing publication&  120&4 &838\\ \hline
    \end{tabular}
    \caption{Research interns who participated in our one-week user study. Research areas, project stage, the time spent completing the SIA template, and the number of revisions were all self reported. Length is the word count of their template responses.}
    \label{tab:interns}
\end{table}

\subsection{Perceived Limitations and Potential Future Iterations of the SIA Template} \label{limitations of SIA}

Despite overall positive feedback, a number of participants (8 out of 15) felt neutral or disagreed (i.e., selected either ``neutral,'' ``disagree,'' or ``strongly disagree'' in the exit survey) with the statement that “I would use this template again in the future” ($\mu$=3.53, $\sigma$=4.27). Through the exit interview, we identified \textbf{two main reasons }why participants were hesitant about using the template. First, the \textbf{current SIA template is viewed as too long}. Around half of the participants (7/15) disagreed that “the template length is suitable to help with self-reflection,” with three selecting ``neutral'' and four selecting “disagree” ($\mu$=3.40, $\sigma$=1.06). T04's comment in the exist survey captured many participants' concerns: \textit{“It was long --- I would do this if properly incentivized (if we were going to talk about it in a meeting for a project, for example), but it's maybe too long for me to fill out voluntarily on my other projects when I already have other tasks to do. On the flip side, it was really useful for thinking about new research directions (that help with the societal impact goal), so I might voluntarily use it.”}  \looseness=-1

Second, the \textbf{current SIA template is not fully suitable for some research areas.} Around half of the participants (7/15)  felt neutral or disagreed with the statement that “I believe the current template is appropriate for my particular area of research” ($\mu$=3.73, $\sigma$=1.10). T02, who works on reinforcement learning theory, shared during the interview that \textit{``most questions in section 3 are not applicable for this project''} because their project set-up is very specific and only a small group of reinforcement learning theorists can make use of their current results. Similarly, T10 shared in their exit survey that: \textit{“I think this template is really well suited to people doing systems design and development work. For those working in more critical/communications/sociology field the template isn't as applicable.”} However, during the exit interview, T10 mentioned that they could see that with a modification, ideally by researchers and domain experts in their own areas, they could \textit{“make the template more useful to qualitative research based on the current structure and content.”}   \looseness=-1

Validating experts' suggestions that the template should help researchers prioritize risks and develop appropriate responses  (DC-6 in Section \ref{DC for template}), many participants, such as T01, T03, T07--T11, and T13--T15, shared in their exit interviews that they found the scaffold for determining the seriousness of possible harms and for planing mitigation useful. 
However, some participants shared that they\textbf{ desired even greater support to effectively evaluate the likelihood and magnitude of harms and to formulate appropriate mitigations}. For example, T08 believed that they needed to engage with users and discuss design alternatives to identify suitable mitigations. This need for additional support explains why participants rated ``Helping me prepare mitigations for the potential negative societal impact of my research'' relatively low ($\mu$=3.67, $\sigma$=1.13). In the exit interviews, some participants, such as T01, T04, and T13, suggested that we make the template more interactive, perhaps using generative AI to aid in brainstorming. \looseness=-1

In line with our observations from the co-design study (DC-2 in Section \ref{DC for template}), around half of the participants (8/15) indicated they navigated back and forth between sections when completing the “Potential Impact” portion of the template. However, participants \textbf{desired more support for non-linear thinking when contemplating the many possible dimensions of societal impact}. For example, T05 
shared in the survey that they \textit{``had difficulty going back and forth between sections (stakeholders, harms, etc)''} and suggested alternative presentation, such as a table that \textit{``detail[s] each consideration (columns) for each stakeholder (rows).''} During the exit interview, many participants discussed the idea of making the template more interactive and dynamic to allow researchers to move between sections more easily.
\looseness=-1

Interestingly, 7 out of 15 participants disagreed that ``the template helped me overcome some of the practical challenges that have kept me from considering and grappling with the potential negative societal impact of my research'' ($\mu$=3.47, $\sigma$=1.19). Through the exit interview, we learned that this was because they \textit{``didn't perceive any practical challenges preventing [them] from assessing the negative societal impact in the first place''} (T09). This feedback may be attributed to the recruitment of industry research interns rather than full-time industry researchers for the one-week user study, a limitation we discuss in the following section.

\section{Discussion}

\subsection{Unique Challenges for Industry Computing Researchers} \label{dis: unique challenge}

Our work extends prior empirical research by highlighting the unique challenges faced by industry computing researchers when grappling with the potential negative societal impact of their work (Section \ref{empirical findings}). In particular, they encounter challenges shaped by team structures and organizational hierarchies (Section \ref{relationship with product}), organizational dynamics and culture (Section \ref{organizational dynamics}), and infrastructure provided --- or not provided --- by their companies (Section \ref{lack infrastructure}).

Like their peers in the academy, industry computing researchers aim to publish and present at peer-reviewed conferences. As a result, they share similar concerns as researchers in the academy who worry that highlighting the potential negative societal impacts of their work might decrease the likelihood of their work being accepted at these conferences --- and thus affect their careers \cite{do2023s, bernstein2021ethics, ada2022looking}. But industry computing researchers differ from their peers in the academy because their attempts to grapple with these impacts may also be at odds with their companies' interests, placing their jobs at more immediate risk \cite[c.f.][]{widder2023open}.  

Our findings also indicate that even when industry computing researchers are encouraged to consider the potential negative impacts of their work, they often face communication and collaboration barriers with product or compliance teams. These barriers arise from differences in knowledge and vocabulary (Section \ref{lack infrastructure}), differing timelines and priorities stemming from different job responsibilities (Section \ref{organizational dynamics}), as well as a lack of transparency into and control over how researchers' work will be incorporated into products and services (Section \ref{relationship with product}). One promising direction is to create ``boundary objects'' that can sit among researchers, product teams, and legal and compliance teams to facilitate collaboration \cite{park2021facilitating, star1989institutional, liao2023designerly, subramonyam2022solving}. The SIA template itself could potentially serve as a prototype for developing future boundary objects to help researchers communicate their findings in a format and language accessible to other team members within their organizations. Future research should also explore how the SIA template can be integrated with other existing responsible computing tools that have been adopted in industry \cite{mitchell2019model, AIX360API, AIF360API, googlePAIR, gebru2021datasheets,  bird2020fairlearn}. \looseness=-1

In line with the extensive empirical research on responsible computing and AI development under the organizational cultures and dynamics within industry settings \cite{holstein2019co, madaio2020co, madaio2022assessing, deng2022exploring, deng2023investigating, deng2023understanding, rakova2021responsible, widder2023s, yildirim2023investigating, wang2023designing, wong2021tactics, passi2018trust, passi2019problem}, we recognize that the challenges faced by industry researchers in addressing negative societal impacts often stem from business-oriented incentives that prioritize existing customers, larger consumer groups, or wealthier markets segments. Addressing these challenges requires political changes that fundamentally alter the prevailing culture of innovation \cite{madaio2022assessing, rakova2021responsible, wong2021tactics} and power dynamics between technology workers and leadership \cite{widder2023dislocated, widder2023s, deng2023investigating, passi2018trust}. While we offer practical approaches that could benefit industry computing researchers and technology companies in the short term, broader changes to the technology industry might be required. \looseness=-1

\subsection{Beyond a Societal Impact Assessment Template} \label{dis: beyond SIA}

Guiding industry computing researchers to engage in self reflection and complete the impact assessment template is merely \textit{an initial step} towards addressing the potential negative impacts from industry research. To be effective, impact assessments conducted within companies must be complemented by external mechanisms and policies to drive meaningful changes in technology design and deployment~\cite{PAI2021managing,ada2022looking, metcalf2021algorithmic}.\looseness=-1

Social science and legal scholars have cautioned against ``performative accountability,'' where companies may appear committed to accountability (e.g., by conducting an impact assessment) but engage in practices that undermine public values in reality \cite{waldman2021industry, gansky2022counterfacctual, edelman2011comply, vigneau2015firms}. Therefore, future researchers, practitioners, and policymakers should draw lessons from the experiences of other fields with impact assessment processes
such as environmental impact assessments \cite{cashmore2004interminable, petts2009handbook}, human right impact assessments \cite{tedaldi2016human}, and data protection impact assessments \cite{PIA_GDPR, PIA_Canada}, among others, in seeking to develop an effective impact assessment for research \cite{andrade2021ai, metcalf2021algorithmic, moss2021assembling, reisman2018algorithmic}. 
Future work should also draw from works in these domains to explore possible ways to try to get people to actually use the impact assessment, ranging from large-scale changes to the norms of the research community to formal policy requirements within particular institutions \cite{rubambiza2024seam}. \looseness=-1

In their recent work, Metcalf et al. have also highlighted the risk of conducting impact assessments without the meaningful engagement from diverse stakeholders, as the assessment outcomes might be ``inappropriately distant from the harms experienced by people'' \cite{metcalf2021algorithmic}. Impact assessments of industry research additionally need to be (1) evaluated by researchers and experts with diverse backgrounds and (2) contested by the communities impacted by the research. Recent work done by Kieslich et al. also demonstrated the importance of engaging end users in anticipating AI impacts \cite{kieslich2023anticipating}. Drawing from prior HCI research  \cite[c.f.][]{shen2021value,wong2021timelines,subramonyam2022solving, park2021facilitating, shen2020designing, shen2022model, smith2020keeping}, future research could explore tools and processes to facilitate meaningful conversations and deliberations between researchers and diverse community members in discussing how certain research might impact their lives. Researchers should also work closely with policymakers to explore the possibility of incorporating these tools into formal regulatory processes (e.g., in the decision-making of government funding bodies) \cite{raji2020closing, metaxa2021auditing, reisman2018algorithmic}. \looseness=-1

Future work should also consider changes to the format of the SIA template to try to overcome some of limitations we identified in Section \ref{limitations of SIA}. Our co-design mainly focused on the content of a static template; future research could explore new forms of impact assessment beyond a template, such as interactive tools or guidelines \cite[c.f.][]{shen2021value, wong2021timelines, TarotCards, nathan2008envisioning, pang2023anticipating, yang2019sketching, hong2021planning, smith2020keeping, pang2024blip}. An increasing line of recent work has also explored the use of emerging generative AI technologies to assist in brainstorming and creating incident cases \cite{buccinca2023aha, rastogi2023supporting, pang2023anticipating, park2022social, kieslich2023anticipating, pang2024blip}. However, these studies have also highlighted the caveats and limitations of using generative AI in tasks like impact assessment \cite{buccinca2023aha, rastogi2023supporting}. More work needs to be done to provide more engaging and effective scaffolding.

\subsection{Limitations}\label{limitations}

As mentioned in Section \ref{limitations of SIA}, one key limitation of our work is the \textbf{validity of using industry research interns from a single company for the one-week user study}. Although testing with research interns still provided valuable insights into the usability and usefulness of our template, their experience and challenges might differ from those of full-time researchers due to variations in their roles, responsibilities, and experiences. In addition, although a one-week study period allowed us to start to understand how researchers would use the template, ideally, we would observe its usage throughout the entire lifecycle of a research project --- from formulating research questions to executing the research, publishing papers, and integrating findings into actual products. To this end, future work should explore ecologically valid methods for evaluating the template within an organizational context. For instance, researchers could conduct a longitudinal study to examine how full-time industry researchers use a version of the SIA template, tailored to their specific research domain and organizational dynamics, at different stages of a research project \cite[c.f.][]{passi2018trust}. Future work could also explore how different research roles in industry (e.g., research interns, full-time researchers, and research managers) might have different incentives when using the impact assessment for their research. Researchers could also collaborate with conference organizers to study how researchers fulfill conference requirements using the SIA template \cite[c.f.][]{stelmakh2021prior}, or perform a between-subjects study comparing the quality of impact statements between researchers who use the template and those who do not \cite[c.f.][]{stelmakh2023large}. \looseness=-1

Another limitation is that our purposive and snowball sampling approach limited \textbf{the diversity of practitioners involved}. Despite efforts to recruit participants from diverse regions, the majority of industry researchers and experts in our study were from the U.S. and U.K. Therefore, future research is necessary to determine whether our empirical findings and design considerations are applicable in non-Western organizational contexts \cite{sturm2015weird, jobin2019global}. Moreover, \textbf{our positionality} as U.S.-based researchers with backgrounds in AI and HCI influenced our data interpretation and the integration of participant feedback into the template’s design and development \cite{holmes2020researcher}. Therefore, we call for future work on this topic conducted by individuals with different backgrounds, including those outside academia and industry.

Finally, ideally, we would have left more time between the interviews and co-design sessions to allow more reflections on the interview data to update the impact assessment template.

\subsection{Potential Negative Societal Impact and a Plan for Mitigation} \label{our societal impact}

\citet{olteanu2023responsible} recently argued that research on ethical issues in computing can also create their own risks of harm. Our work is no exception. To assess these risks, we conducted an impact assessment of our own work using the SIA template. We identified a number of dangers. First, researchers and companies who use our template might be left with a \textbf{false sense of confidence} after completing their impact assessments, mistakenly believing that they have exhausted all potential harms that might arise due to the research in question. Second, they might also conclude that it is enough to \textbf{just conduct the assessment without taken meaningful action} towards addressing any of the identified harms. To address these risks, we explicitly highlight them within the template itself, in line with \ref{DC for template}, and  we offer concrete future steps for other researchers and organizations to avoid this in Section \ref{dis: beyond SIA}. When open-sourcing the SIA template through a project web page, we plan to further underscore the appropriate use of impact assessments and the importance of continuous monitoring. \looseness=-1

\textbf{For-profit companies might exploit the SIA template as a tool for ethics washing} \cite{ali2023walking, deng2023understanding, widder2023dislocated, rakova2021responsible}, as others have observed in prior research studying responsible AI \cite{deng2023understanding, rakova2021responsible, madaio2020co}, privacy and data protection \cite{waldman2021industry, tahaei2021privacy, gutfleisch2022does, bamberger2011privacy}, and environmental justice \cite{Miceli2021ThrivingNJ,petts2009handbook, brown1998making, vigneau2015firms}. Prior work on algorithmic auditing \cite{do2023s, prunkl2021institutionalizing, bernstein2021ethics, metaxa2021auditing, diakopoulos2015algorithmic, bandy2020auditing} offers helpful suggestions for how to ensure the integrity of the impact assessment process and bring about meaningful change, often via external review. \looseness=-1

Finally, conducting an impact assessment might  create an unreasonable \textbf{burden on researchers}, in line with feedback from the user study. Like IRBs, impact assessments also  run the risk of stifling research freedom if the procedures are designed poorly \cite{silberman2011burdens, grady2015institutional}. To guard against these risks, organizations should formally recognize the ``invisible labor'' that goes into responsible computing practices  \cite[cf.][]{deng2023investigating, wong2021tactics, star1999layers, ali2023walking}, appropriately accounting for these efforts in employees' job descriptions and performance reviews. Organizations should also establish and support teams that can train researchers, provide advice, and help reason through difficult questions \cite[cf.][]{bernstein2021ethics, PAI2021managing}. \looseness=-1

\section{Conclusion}

Through interviews with 25 industry computing researchers, co-design sessions with both these researchers and 16 impact assessment experts, and a user study with 15 industry research interns, this paper (1) sheds light on current perceptions, practices, and challenges among computing researchers seeking to confront the potential negative societal impacts of their work (Section \ref{empirical findings}), (2) provides guidance for researchers, organizations, and policymakers to improve these practices and overcome identified challenges (Section \ref{design considerations}), and (3) offers a version of the SIA template with tangible scaffolding materials (Section \ref{SIA template}) along with its user study (Section \ref{user study results}) to help computing as a field of research do more to grapple with the consequences of its work. By connecting and contrasting these contributions with literature from CSCW, design, and the broader fields of HCI and social science, we illuminate paths for future researchers, industry practitioners and organizations, and policy makers to better support responsible research practices in industry settings and beyond. \looseness=-1


\begin{acks}
We thank all participating industry researchers, impact assessment experts, and industry research interns for making this work possible. We are grateful to Snehal Prabhudesai, Yuda Song, and Samantha Dalal for joining the pilot interview study. We thank Ken Holstein, Motahhare Eslami, Kevin Feng, and Ezra Awumey for their feedback on the draft. Finally, we thank Alicia Edelman Pelton, Forough Poursabzi-Sangdeh, Sunnie Kim, Rock Pang, Michael Madaio, Nari Johnson, Hoda Heidari, Alex London, Jason Hong, Jeff Bigham, Lauren Wilcox, Eric Horvitz, the CMU FEAT reading group, the Microsoft Research FATE group, and Microsoft's Aether RADD focus group for many useful discussions. \looseness=-1
\end{acks}

\bibliographystyle{ACM-Reference-Format}
\bibliography{citation}

\appendix

\section{Appendix}

\subsection{Evolution of the SIA Template}

In the supplementary material, we include the initial prototype of the SIA template (\textbf{V1, used with participants P01--P05}) and the version used in the user study (V7).  Here we briefly describe the major changes that were made in each version of the template between V1 and V7. In every iteration, we additionally made wording and formatting changes throughout the template based on participants' feedback.
\\

\noindent\textbf{V2 (used with participants P06--P11)}:

\begin{itemize}
    \item Added more examples in the scaffolding provided in the ``Potential Negative Societal Impact'' section to cover a more diverse range of computing research topics.
    \item Moved the ``Misuse (unintentional, accidental misuse)'' section right after the ``Known Limitation'' section within the ``Potential Negative Societal Impact'' section since participants P01--P05 often naturally thought about misuse after considering limitations.
    \item Rephrased the questions in the first two sections (``Research Project Background'' and ``Intended Users and Applications'') to make them more directly applicable to research projects instead of product and service development.
\end{itemize}

\noindent\textbf{V3 (used with participants P12--P15)}:

\begin{itemize}
    \item Emphasized the non-linear nature of brainstorming stakeholders by reminding researchers to revisit the stakeholders section if they think of more stakeholders later.
    \item Added examples to the scaffolding in the ``Intended Positive Societal Impact'' section, as participants often needed help in this area.
    \item Added a sentence to highlight the differences between assessing societal impact and preventing risks to human subjects, as participants often confused these concepts.
\end{itemize}

\noindent\textbf{V4 (used with participants P16--P18 and E01--E03)}:

\begin{itemize}
    \item Further emphasized the heterogeneous, non-linear thinking processes of impact assessment by adding notes suggesting researchers complete the sections of the template non-sequentially, in their preferred order.
    \item Added a section titled "Why would you want to use this template" at the top as motivation.
    \item Added a 3-minute introduction video as onboarding material, as described in Section \ref{SIA template}.
\end{itemize}

\noindent\textbf{V5 (used with participants P19--P22 and E04--E10)}:

\begin{itemize}
    \item Reorganized the categories within the stakeholders section to distinguish industry, civil society, and governments, with a concise description of each to scaffold brainstorming, as suggested by impact assessment experts.
    \item Added illustrations (e.g., Figures \ref{fig:SIA} and \ref{fig:research}) throughout the template to help researchers better understand the organization of the template and key concepts.
    \item Added more categories of limitations in the ``Known Limitations'' section to cover potential limitations of design and qualitative industry computing research.
    \item Provided a keyword summary for all scaffolding examples throughout the template for more efficient access.
\end{itemize}

\noindent\textbf{V6 (used with participants P23--P25 and E11--E16)}:

\begin{itemize}
    \item Added instructions to remind researchers to document different versions of their project, their thinking process, and the challenges they encountered. Encouraged transparency about struggles throughout the template.
    \item Incorporated the sentence ``While thinking about the scenarios, consider which group(s) of stakeholders you identified in Section 3 might be harmed. In what ways?'' into all questions in the ``Potential Impact'' section to better facilitate thinking about negative impact and stakeholders together. \looseness=-1
    \item Highlighted the limitations of the impact assessment template, noting that it is not meant to replace other existing mechanisms or compliance processes in the company, within the onboarding video and instructions.
    \item Reorganized the strategies for addressing potential negative societal impact based on feedback from impact assessment experts.
\end{itemize}

\noindent\textbf{V7 (used with participants T01--T15)}:

\begin{itemize}
    \item Based on feedback from impact assessment experts, expanded the ``Planning for Mitigations'' section to guide researchers in determining the seriousness of the risks posed by potential negative societal impacts and how to prioritize them. Added scaffolding questions to prompt researchers to consider the likelihood, magnitude, concentration, and time horizon of the potential negative societal impact.
    \item Further improved the clarity of the questions and the transitions between all sections.
        \item Streamlined the template to ensure the format and structure were ready for the user study.
\end{itemize}

\end{document}